\documentclass{article}
\usepackage{epsfig}
\usepackage{psfrag}
\usepackage{subfigure}
\usepackage{amsmath,amssymb,amscd,amsthm}
\usepackage{upgreek}
\usepackage{color}
\newtheorem{rem}{Remark}[section]
\newcommand{\br}{\begin{rem}}
\newcommand{\er}{\end{rem}}
\newtheorem{ex}[rem]{Example}
\newcommand{\bex}{\begin{ex}}
\newcommand{\eex}{\end{ex}}
\newtheorem{Def}[rem]{Definition}
\newcommand{\bd}{\begin{Def}}
\newcommand{\ed}{\end{Def}}
\newtheorem{theorem}[rem]{Theorem}
\newcommand{\bt}{\begin{theorem}}
\newcommand{\et}{\end{theorem}}
\newtheorem{lemma}[rem]{Lemma}
\newcommand{\bl}{\begin{lemma}}
\newcommand{\el}{\end{lemma}}
\newtheorem{cor}[rem]{Corollary}
\newcommand{\bc}{\begin{cor}}
\newcommand{\ec}{\end{cor}}
\newtheorem{con}[rem]{Conjecture}
\newcommand{\bcon}{\begin{con}}
\newcommand{\econ}{\end{con}}
\newtheorem{prop}[rem]{Proposition}
\newcommand{\bp}{\begin{prop}}
\newcommand{\ep}{\end{prop}}
\newcommand{\be}{\begin{equation}}
\newcommand{\ee}{\end{equation}}
\newcommand{\bea}{\begin{eqnarray}}
\newcommand{\eea}{\end{eqnarray}}
\newcommand{\pa}{\partial}
\newcommand{\nn}{\nonumber}
\newcommand{\adots}{\mathinner{\mkern2mu\raise1pt\hbox{.}\mkern2mu
\raise4pt\hbox{.}\mkern2mu\raise7pt\hbox{.}\mkern1mu}}

\newcommand{\RA}{{\mathrm A}}
\newcommand{\RB}{{\mathrm B}}
\newcommand{\RC}{{\mathrm C}}
\newcommand{\RD}{{\mathrm D}}

\newcommand{\CH}{{\cal H}}

\newcommand{\Z}{{\mathbb Z}}
\newcommand{\C}{{\mathbb C}}

\newenvironment{prf}{\trivlist \item [\hskip
\labelsep {\bf Proof:}]\ignorespaces}{\qed \endtrivlist}

\title{Periodic Cluster Mutations and Related Integrable Maps}
\author{Allan P. Fordy ,  \\ School of Mathematics, \\
University of Leeds. \\ Leeds LS2 9JT, UK.\\
e-mail: a.p.fordy@leeds.ac.uk}

\begin{document}

\maketitle

\begin{abstract}
One of the remarkable properties of cluster algebras is that any cluster, obtained from a sequence of mutations from an initial cluster, can be written as a Laurent polynomial in the initial cluster (known as the ``Laurent phenomenon'').  There are many nonlinear recurrences which exhibit the Laurent phenomenon and thus unexpectedly generate integer sequences.  The mutation of a typical quiver will not generate a recurrence, but rather an erratic sequence of exchange relations.  How do we ``design'' a quiver which gives rise to a given recurrence?  A key role is played by the concept of ``periodic cluster mutation'', introduced in 2009.

Each recurrence corresponds to a finite dimensional map.  In the context of cluster mutations, these are called ``cluster maps''.  What properties do cluster maps have?  Are they integrable in some standard sense?

In this review I describe how integrable maps arise in the context of cluster mutations.  I first explain the concept of ``periodic cluster mutation'', giving some classification results.  I then give a review of what is meant by an integrable map and apply this to cluster maps.
Two classes of integrable maps are related to interesting monodromy problems, which generate interesting Poisson algebras of functions, used to prove complete integrability and a linearisation.  A connections to the Hirota-Miwa equation is explained.
\end{abstract}

\tableofcontents

\section{Introduction}

The main purpose of this review is to explain how {\em integrable maps} arise in
the context of cluster mutations.  These maps will be derived from recurrences, so we must first ask how {\em recurrences} arise in
the context of cluster mutations.  To be specific, consider an $N^{th}$ order recurrence in a single variable $x_n$, of the form
\be\label{GenRec}
x_n x_{n+N}  = F(x_{n+1}, \ldots ,  x_{n+N-1}),
\ee
where $F$ is a polynomial in $N-1$ variables.  How can we represent such a recurrence as a sequence of cluster mutations?

Recall that cluster exchange relations (satisfied by the generators of a cluster algebra) are derived by looking at the nodes of an associated {\em quiver} (see Section \ref{quiver}).  Each {\em quiver mutation} leads to a specific exchange relation, but also to a change of the quiver.  If we randomly choose even a fairly simple quiver, a little experimentation (using Keller's Java Applet \cite{keller}, for example) shows that mutation {\em typically} generates a series of completely different quivers with a wild growth of the number of arrows connecting nodes.  Therefore, each successive exchange relation will involve a {\em different} formula for $x_k\mapsto \tilde x_k$, so cannot possibly lead to a regular formula such as (\ref{GenRec}).  We need to construct a quiver with a special property that after mutation at node $k$ (say), at least one node of the new quiver should have the same configuration of arrows as at node $k$ of the original quiver.  After playing with Keller's Java Applet, this seems like a tall order!  Nevertheless, in \cite{f11-2} we introduced the concept of ``periodic cluster mutation'', which gives a way of building quivers for which a
{\em specific sequence of mutations} leads to a recurrence. One of the purposes of this paper is to explain this construction.

Now, supposing we have a way of generating a recurrence from a sequence of cluster mutations, what properties can we expect this recurrence to have?
One of the remarkable properties of cluster algebras \cite{02-3} is that any cluster, obtained from a sequence of mutations from an initial cluster, can be written as a Laurent polynomial in the initial cluster (known as the ``Laurent phenomenon'').  In particular, if $\{x_n\}_{n=1}^\infty$ is a sequence generated by cluster mutations, we expect it to possess the Laurent property: ie that the general term $x_n$ can be written as a Laurent polynomial in the initial conditions, meaning a polynomial in $\{x_n^{\pm 1}\}_{n=1}^N$.  This property cannot be expected to hold for a sequence generated by (\ref{GenRec}) for arbitrary $F$ since, for example, the calculation of $x_{2N+1}$ requires division by the polynomial $F(x_2,\dots ,x_N)$.  Hence we generally expect $x_n$ to be {\em rational}, with nontrivial {\em polynomials} in the denominators. This means that miraculous cancellations must take place, rendering the denominator a \underline{monomial}!  An immediate consequence of the Laurent property is that if we take initial conditions $x_1=\cdots =x_N=1$, then the recurrence generates a sequence of {\em integers}.

The archetypal example is the Somos-$4$ recurrence
\be\label{s4-sequ}  %
x_n x_{n+4} = x_{n+1}x_{n+3}+x_{n+2}^2,
\ee  %
which (with initial conditions $x_1=x_2=x_3=x_4=1$) generates the {\em integer sequence}:
$$
1,1,1,1,2,3,7,23,59,314,1529,8209,\dots
$$
(see ``The On-Line Encyclopedia of Integer Sequences'' \cite{sloane} and \cite{91-17}).  Other examples are the Somos-$5$ and $6$ recurrences
\bea  %
&& x_n x_{n+5}=x_{n+1}x_{n+4}+x_{n+2}x_{n+3},  \label{s5-sequ}  \\
&&  x_n x_{n+6} = x_{n+1} x_{n+5}+x_{n+2} x_{n+4}+x_{n+3}^2 .  \label{s6-sequ}
\eea  %
In fact we can write a ``Somos-$N$ recurrence'' for all $N$, but from Somos-$8$ onwards, they no longer possess the Laurent property.  One might be tempted to think that any recurrence with the Laurent property should be derivable through cluster mutation, but the general cluster exchange relation (see (\ref{ex-rel})) has only {\em two terms} on the ``right hand side'', so (\ref{s6-sequ}) {\em cannot} be so obtained.  Nevertheless, Fomin and Zelevinsky did introduce other cluster-related techniques in \cite{02-2}, with which they could prove the Laurent property for this more general class of recurrence.  Since both (\ref{s4-sequ}) and (\ref{s5-sequ}) have the {\em correct} two-term form, we may hope that they are connected to some cluster mutations.  Indeed, quivers corresponding to these were derived in \cite{f11-2} (see Section \ref{quiver}).

Given the recurrence (\ref{GenRec}), we can consider a corresponding {\em cluster map} $\varphi:\C^N\rightarrow \C^N$ of the form
\be\label{GenMap}  %
\varphi:(x_1,\dots ,x_N)\mapsto (x_2,\dots ,x_N,x_{N+1}),\quad\mbox{where}\;\;\; x_{N+1}=\frac{F(x_2,\dots ,x_{N})}{x_1},
\ee  %
which defines a {\em discrete dynamical system} on $\C^N$.

What properties (other than being Laurent) can we expect this map to possess?  It is known \cite{07-2} that both (\ref{s4-sequ}) and (\ref{s5-sequ}) are related to special cases of the QRT map \cite{88-5}, a well known $18$ parameter family of integrable maps of the plane.  Again, we might be tempted to think that {\em all} maps that are derived through cluster mutation are integrable.  However, numerical experimentation, or applying one of the ``tests for integrability'' (see Section \ref{entropy}) soon shows that most cluster maps are {\em non-integrable}.  We must, therefore, address the question of {\em isolating} integrable cases and possibly giving some sort of classification.  Such ``tests for integrability'' are, in fact, only an indication, usually just supplying {\em necessary} conditions.

When we say a map is integrable, we mean {\em integrable in the Liouville sense}.  This means that we have a usual completely integrable Hamiltonian system, whose commuting Hamiltonian flows are {\em continuous symmetries} of our discrete dynamical system.  Our maps (perhaps after reduction to a lower dimensional manifold) will be {\em Poisson maps} (generalisations of {\em canonical transformations}), subject to some additional invariance properties.  To keep this review self-contained, Section \ref{poisson} gives a brief introduction to these concepts.

Given a map, it is a nontrivial task to find an appropriate symplectic or Poisson structure which is invariant under its action.  Remarkably, in the case of cluster maps derived from periodic quiver mutation, this problem can be solved algorithmically.  An invariant (pre-)symplectic structure is built out of the matrix $B$ which defines the quiver.  On symplectic leaves, this in turn defines an invariant Poisson bracket (see Section \ref{symplectic}).  However, there is {\em no} algorithm for finding the appropriate number of invariant, Poisson commuting functions.  Nevertheless, for some classes of cluster maps it is possible to find these.

In Section \ref{primit} we consider a special class of recurrence related to the simplest period $1$ quivers (the ``primitives'').  For an {\em even} number of nodes, it was shown in \cite{f11-1} these are {\em bi-Hamiltonian} and this can be used to algorithmically build the commuting integrals. Later, it was shown in \cite{f14-1} that for any number of nodes it is possible to build a $2\times 2$ matrix monodromy problem, from which we can derive the commuting integrals.  It is also possible to show that the sequence $\{x_n\}_{n=1}^\infty$ generated by the nonlinear ``primitive'' recurrences, satisfies a higher order {\em linear} recurrence.  Similar results can be derived for another recurrence, described in Section \ref{pert}, only this time the monodromy matrix is $3\times 3$.  For this class it is sometimes (but not always!) possible to find the appropriate number of invariant, commuting integrals to prove complete integrability.

The Somos-$4$ and Somos-$5$ recurrences cannot be linearised, but their corresponding maps do have a Lax representation.  This follows from the fact that these recurrences (and a larger class, containing them) are reductions of the Hirota-Miwa (discrete KP) Equation.  This is briefly described in Section \ref{grmaps}.

This paper is mainly a review of  \cite{f11-2,f11-1,f11-3,f14-1}, but in the conclusions I mention some recent developments and discuss open problems.

\section{Quiver Mutation}\label{quiver}
\setcounter{equation}{0}

A quiver is a {\em directed graph}, consisting of $N$ nodes with directed edges
between them.  There may be several arrows between a given pair of vertices,
but for cluster algebras there should be no $1$-cycles (an arrow which starts
and ends at the same node) or $2$-cycles (an arrow from node $a$ to node $b$,
followed by one from node $b$ to node $a$). A quiver $Q$, with $N$ nodes, can
be identified with the unique skew-symmetric $N\times N$ matrix $B_Q$ with
$(B_{Q})_{ij}$ given by the number of arrows from $i$ to $j$ (with the obvious
sign convention).

\bd[Quiver Mutation]\label{d:mutate}   %
Given a quiver $Q$ we can mutate at any of its nodes. The mutation of $Q$ at
node $k$, denoted by $\mu_k Q$, is constructed (from $Q$) as follows:
\begin{enumerate}
\item  Reverse all arrows which either originate or terminate at node $k$.
\item  Suppose that there are $p$ arrows from node $i$ to node $k$ and $q$
arrows from node $k$ to node $j$ (in $Q$). Add $pq$ arrows going from node $i$
to node $j$ to any arrows already there. \item Remove (both arrows of) any
two-cycles created in the previous steps.
\end{enumerate}
\ed    %
Note that in Step $2$, $pq$ is just the number of paths of length $2$ between
nodes $i$ and $j$ which pass through node $k$.

Quiver mutation is very visual and easy to understand, but for computations
we use the following:
\bd[Matrix Mutation]   %
Let $B$ and $\tilde B=\mu_k B$ be the skew-symmetric matrices corresponding to the
quivers $Q$ and $\tilde Q=\mu_k Q$.  Let $b_{ij}$ and $\tilde b_{ij}$ be the
corresponding matrix entries.  Then quiver mutation amounts to the following
formula
\be   \label{gen-mut}  %
\tilde b_{ij}= \left\{ \begin{array}{ll}
                        -b_{ij} & \mbox{if}\;\; i=k\;\;\mbox{or}\;\; j=k, \\
                        b_{ij}+\frac{1}{2} (|b_{ik}|b_{kj}+b_{ik}|b_{kj}|)
                        & otherwise.
                        \end{array}  \right.
\ee    %
\ed      %

\subsection{Cluster Exchange Relations}

Given a quiver (with $N$ nodes), we attach a variable at each node, labelled
$(x_1,\dots ,x_N)$ (the initial cluster).  When we mutate the quiver we change the associated matrix
according to formula (\ref{gen-mut}) and, {\em in addition}, we transform the
cluster variables $(x_1,\dots ,x_N)\mapsto (x_1,\dots ,\tilde x_k,\dots ,x_N)$, where
\be  \label{ex-rel}   %
x_k \tilde x_k = \prod_{b_{ik}>0} x_i^{b_{ik}}+
            \prod_{b_{ik}<0} x_i^{-b_{ik}},
           \qquad   \tilde x_i = x_i \;\;\mbox{for}\;\; i\neq k .
\ee   %
If one of these products is empty (which occurs when all $b_{ik}$ have the
same sign) then it is replaced by the number $1$.  This formula is called the
(cluster) {\em exchange relation}.  Notice that it just depends upon the
$k^{th}$ column of the matrix.  Since the matrix is skew-symmetric, the
variable $x_k$ {\bf does not} occur on the right side of (\ref{ex-rel}).

After this process we have a new quiver $\mu_k Q$, with a new matrix $\mu_k B$.
This new quiver has cluster variables $(\tilde x_1,\dots ,\tilde x_N)$.
However, since the exchange relation (\ref{ex-rel}) acts as the identity on all
except one variable, we write these new cluster variables as $(x_1,\dots ,\tilde x_k,\dots ,x_N)$.
We can now repeat this process and mutate
$\mu_k Q$ at node $p$ and produce a third quiver $\mu_p\mu_k Q$, with
cluster variables $(x_1,\dots ,\tilde x_k,\dots ,\tilde x_p,\dots ,x_N)$,
with $\tilde x_p$ being given by an analogous formula (\ref{ex-rel}).

\br[Involutive Property]   %
If $p=k$, then $\mu_k^2 Q=Q$, so we insist that $p\neq k$.
\er   %

\subsection{Periodicity}

Before introducing the general concept of periodicity \cite{f11-2}, I give a motivating example, which also illustrates the
above definitions of mutation.  This is the quiver which generates the Somos-$4$ recurrence.

\bex[The Somos-$4$ Quiver $S_4$]  \label{s4exchange}  {\em  %
Consider the quiver (and its associated matrix) of Figure \ref{s4quivermatrix}.
\begin{figure}[hbt]
\begin{minipage}[c]{5cm}
\centering
\includegraphics[width=3cm]{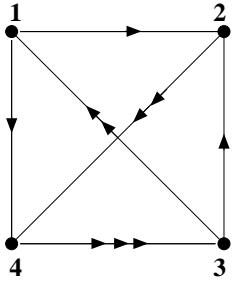}
\end{minipage}
\qquad
\begin{minipage}[c]{5cm}
$$
B= \left(
\begin{array}{cccc}
  0 & -1 & 2 & -1 \\
 1 & 0 & -3 & 2 \\
  -2 & 3 & 0 & -1 \\
  1 & -2 & 1 & 0 \\
\end{array}\right)
$$
\end{minipage}
\caption{The Somos-$4$ quiver $S_4$ and its matrix.}\label{s4quivermatrix}
\end{figure}
The mutation at node $1$ leads to the quiver (and associated matrix) of Figure \ref{s4quiverbmatrix}.
\begin{figure}[hbt]
\begin{minipage}[c]{5cm}
\centering
\includegraphics[width=3cm]{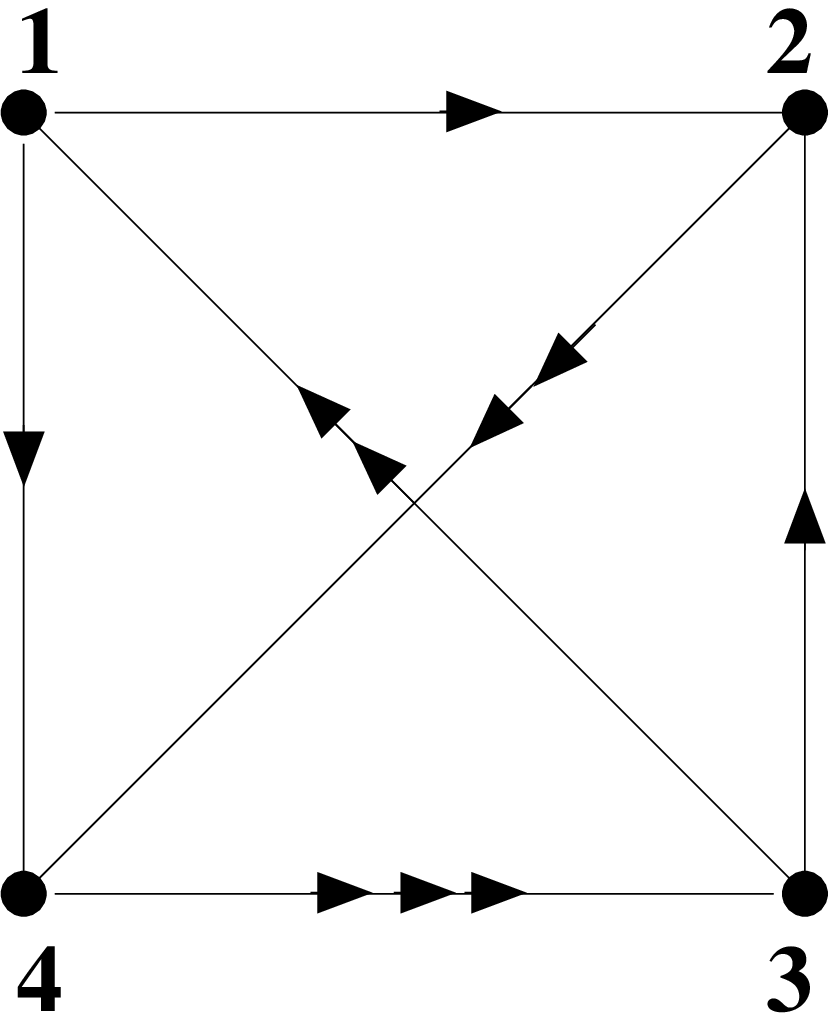}
\end{minipage}
\quad
\begin{minipage}[c]{4cm}
$$
\tilde B= \left(
\begin{array}{cccc}
  0 & 1 & -2 & 1 \\
 -1 & 0 & -1 & 2 \\
  2 & 1 & 0 & -3 \\
  -1 & -2 & 3 & 0 \\
\end{array}\right)
$$
\end{minipage}
\caption{Mutation $\tilde S_4=\mu_1 S_4$ of the quiver $S_4$ at $1$ and its
matrix.}\label{s4quiverbmatrix}
\end{figure}
Placing $x_1,x_2,x_3,x_4$ respectively at nodes $1$ to $4$ of quiver $S_4$
gives the {\em initial cluster}.  Along with the {\em quiver} mutation,
we also have the {\em exchange relation}
\be\label{x1x1t}  %
x_1\tilde x_1 = x_2 x_4+x_3^2.
\ee  %
This corresponds to one arrow coming {\bf into} node $1$ from each of nodes $2$
and $4$ with $2$ arrows going {\bf out} to node $3$.

We can now consider mutations of quiver $\tilde S_4=\mu_1 S_4$.  To avoid too
many ``tildes'', let us write $\tilde x_1 = x_5$, so quiver $\tilde S_4$ has
$x_5,x_2,x_3,x_4$ respectively at nodes $1$ to $4$.  Mutation at node $1$ would
just take us back to quiver $S_4$ (as noted in the above remark).  We compare
the exchange relations we would obtain by mutating at nodes $2$ or $3$.

Mutation at node $2$ would lead to exchange relation
\be\label{x2x2t}  %
x_2\tilde x_2 = x_3 x_5+x_4^2,
\ee  %
whilst that at node $3$ would lead to
\be\label{x3x3t}  %
x_3\tilde x_3 = x_2 x_5^2+x_4^3.
\ee  %
We see that the right hand sides of formula (\ref{x1x1t}) and (\ref{x2x2t}) are
related by a {\bf shift}, whilst formula (\ref{x3x3t}) is entirely {\em
different}.  In fact, it can be seen in Figures \ref{s4quivermatrix} and
\ref{s4quiverbmatrix} that the configuration of arrows at node $2$ of quiver
$\tilde S_4$ is {\em exactly the same} as that at node $1$ of quiver $S_4$,
thus giving the {\em same} exchange relation.  In fact, we have more.  The {\em
whole quiver} $\tilde S_4$ is obtained from $S_4$ by just {\bf rotating the
arrows}, whilst keeping the nodes fixed.  It follows that mutation of quiver
$\tilde S_4$ {\em at node $2$} just leads to a further rotation, with node $3$
inheriting this same configuration of arrows.  If at each step we relabel
$\tilde x_n$ as $x_{n+4}$, the $n^{th}$ exchange relation can be written
\be\label{somos4}  %
x_n x_{n+4} = x_{n+1} x_{n+3}+x_{n+2}^2.
\ee  %
This rotational property of the quiver has lead to a {\bf recurrence}, which in
this case is just Somos-$4$.
}\eex  %

The essential property of the quiver $S_4$ was that mutation at node $1$ just gave
us a copy of $S_4$ {\em up to a permutation} of the indices.  If we insist
(for an $N-$node quiver) that this permutation has order $N$, then (by a change of
labelling) we can represent it by the $N\times N$ matrix
$$
\rho =  \left( \begin{array}{cccc}
                   0 & \cdots & \cdots & 1 \\
                  1 & 0 & & \vdots \\
                     &\ddots & \ddots & \vdots \\
                      & & 1 & 0
                      \end{array} \right),
$$
which acts on the matrix $B$ by
$$
B \mapsto \rho B \rho^{-1}.
$$
In terms of the quiver we write this as $Q\mapsto \rho Q$.

Consider a quiver $Q=Q(1)$, with $N$ nodes, and a sequence of
mutations, starting at node $1$, followed by node $2$, and so on. Mutation at
node $1$ of a quiver $Q(1)$ will produce a second quiver $Q(2)$.  The mutation
at node $2$ will therefore be of quiver $Q(2)$, giving rise to quiver $Q(3)$
and so on. In \cite{f11-2} we defined a {\em period $m$ quiver} as follows.
\bd   %
A quiver $Q$ has \textit{period $m$} if it satisfies $Q(m+1)=\rho^m Q(1)$ (with
$m$ the minimum such integer). The mutation sequence is depicted by
\be   \label{periodchain}   %
Q=Q(1) \stackrel{\mu_1}{\longrightarrow} Q(2) \stackrel{\mu_2}{\longrightarrow}
  \cdots \stackrel{\mu_{m-1}}{\longrightarrow} Q(m)
       \stackrel{\mu_m}{\longrightarrow} Q(m+1)=\rho^m Q(1),
\ee    %
and called the {\em periodic chain} associated to $Q$.

The corresponding matrices would then satisfy $B(m+1)=\rho^m B(1)\rho^{-m}$.
\ed  %

\br[The Sequence of Mutations]  %
We must perform the {\bf correct} sequence of mutations.  For
instance, if we mutate $\mu_1 S_4$ at node $3$, we obtain a quiver which has
$5$ arrows from node $4$ to node $1$, which {\bf cannot} be permutation
equivalent to $Q(1)=S_4$.  As we previously saw, the corresponding exchange
relation (see (\ref{x3x3t})) was also different.
\er  %

\br[Other Definitions of Periodicity]  %
Apparently, examples of periodicity were known in the context of T- and Y-systems, so this concept was formalised in \cite{11-2}.
Indeed, in this paper Nakanishi also defines and develops the theory of periodic cluster mutation, but his motivation is rather different from ours.
\er  %

\subsection{Period $1$ Quivers}

Period $1$ quivers can be completely classified in terms of a special class of quivers,
called primitives.  In our classification, the primitives are the 'atoms' out of which we build the
general period 1 quiver for each N.

\bd[Period $1$ sink-type quivers]\label{d:period1}
A quiver $Q$ is said to be a \emph{period $1$ sink-type quiver}
if it is of period $1$ and node $1$ of $Q$ is a sink.
\ed

\bd[Skew-rotation] \label{d:skewrotation}
We shall refer to the matrix
$$
\tau = \left(\begin{array}{cccc}
                   0 & \cdots & \cdots & -1 \\
                  1 & 0 & & \vdots \\
                     &\ddots & \ddots & \vdots \\
                      & & 1 & 0
                      \end{array} \right).
$$
as a {\em skew-rotation}.
\ed

If node $1$ of $Q$ is a sink, there are no paths of length $2$ through it, and the second part of Definition
\ref{d:mutate} is void.  Period $1$ mutation then reduces to a simple conjugation.

\bl[Period $1$ sink-type equation]\label{l:period1sink}
A quiver $Q$ with a sink at $1$ is period $1$ if and only if
$\tau B_Q \tau^{-1}=B_Q$.
\el

The map $M\mapsto \tau M \tau^{-1}$ simultaneously cyclically permutes the
rows and columns of $M$ (up to a sign), while $\tau^N=-I_N$,
hence $\tau$ has order $N$.
This gives us a method for building period $1$ matrices: we sum over
$\tau$-orbits.

\paragraph{The period $1$ primitives $P_N^{(k)}$.}
We consider a quiver with just a single arrow from node $N-k+1$ to $1$,
represented by the skew-symmetric matrix $R_N^{(k)}$ with
$(R_N^{(k)})_{N-k+1,1}=1, (R_N^{(k)})_{1,N-k+1}=-1$ and
$(R_N^{(k)})_{ij}=0$ otherwise.

We remark that the geometric action of $\tau$ in the above sum is to rotate
the arrow clockwise without change of orientation, except that when the tail
of the arrow ends up at node $1$ it is reversed. It follows that $1$
is a sink in the resulting quiver. Since it is a sum over a $\tau$-orbit,
we have $\tau B_N^{(k)} \tau^{-1} = B_N^{(k)}$, and thus that $P_N^{(k)}$
is a period $1$ sink-type quiver. In fact, we have the simple description:
\be\label{bnk}  %
B_N^{(k)}= \left\{ \begin{array}{ll}
              \tau^k-(\tau^{t})^k, & \mbox{if\ }N=2r+1
                 \mbox{\ and\ }1\leq k\leq r, \mbox{\ or\ }
                           N=2r\mbox{\ and\ }1\leq k\leq r-1, \\
                \tau^r , & \mbox{if\ }N=2r\mbox{\ and\ } k=r,
                \end{array}\right.
\ee  %
where $\tau^{t}$ denotes the transpose of $\tau$.  The choice $1\leq k\leq r$ is
because $R_N^{(k)}$ and $R_N^{(N-k)}$ are in the same orbit.

\begin{figure}[htb]
\centering \subfigure[$P_4^{(1)}$]{
\includegraphics[width=2.5cm]{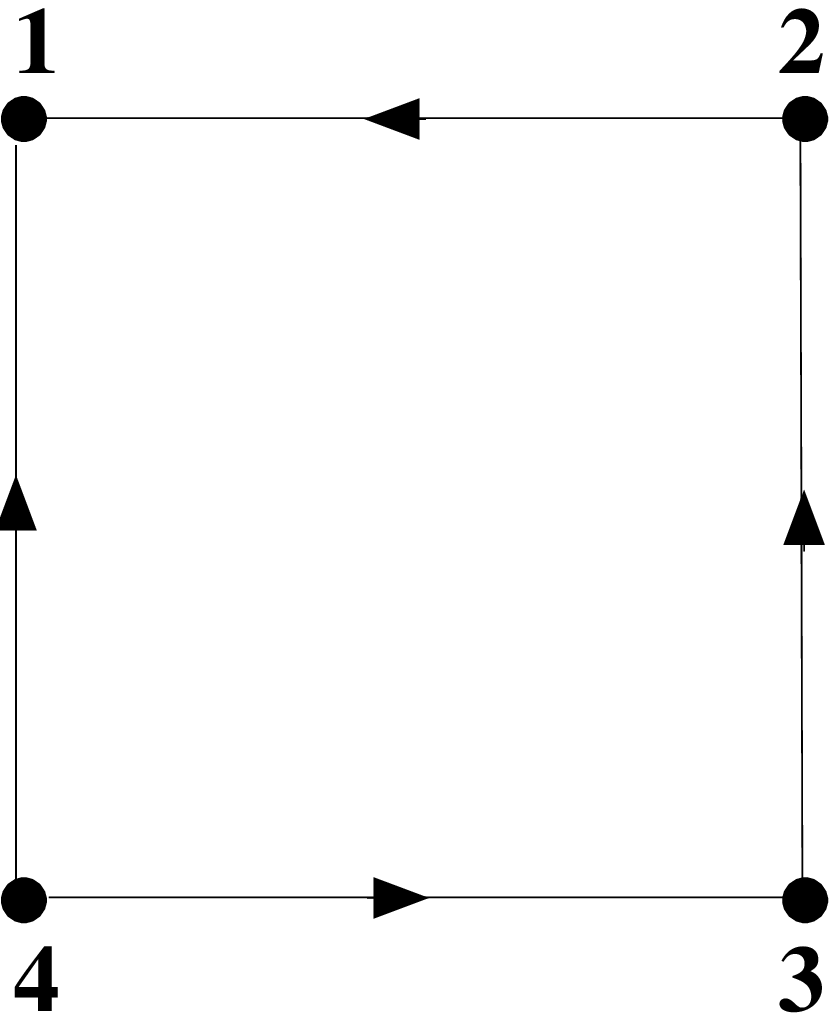}\label{subfig:P41}
}\qquad \subfigure[$P_4^{(2)}$]{
\includegraphics[width=2.5cm]{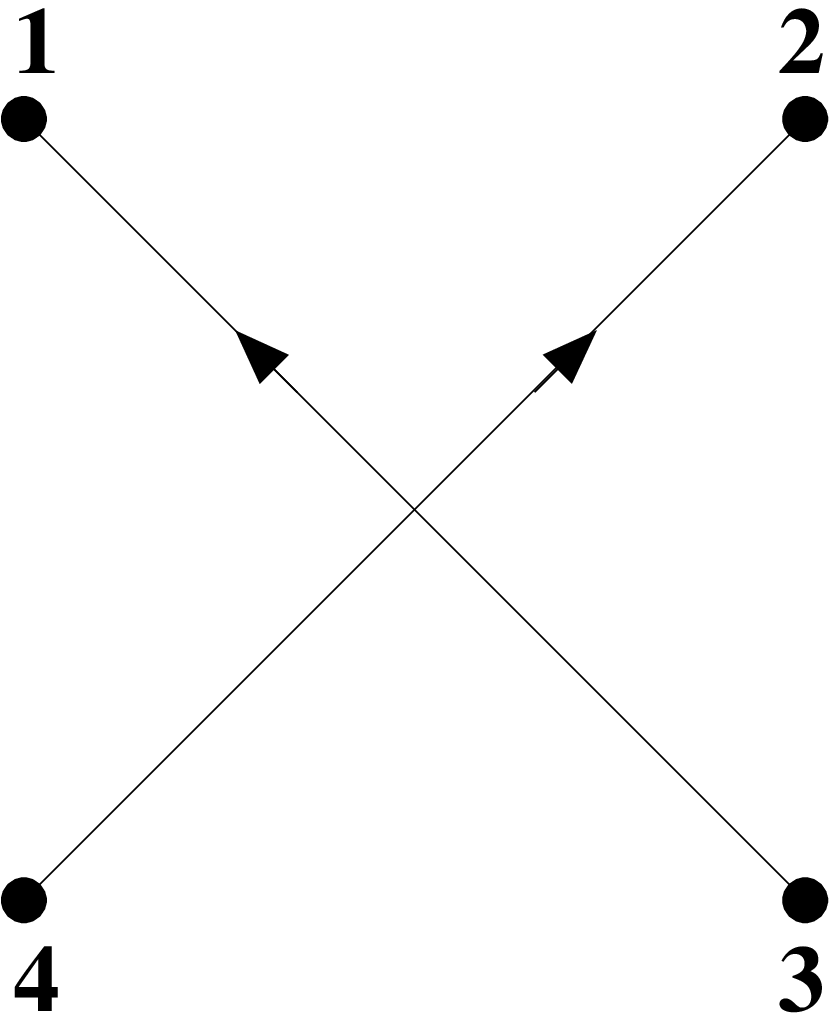}\label{subfig:P42}
}\qquad \subfigure[$P_5^{(1)}$]{
\includegraphics[width=3cm]{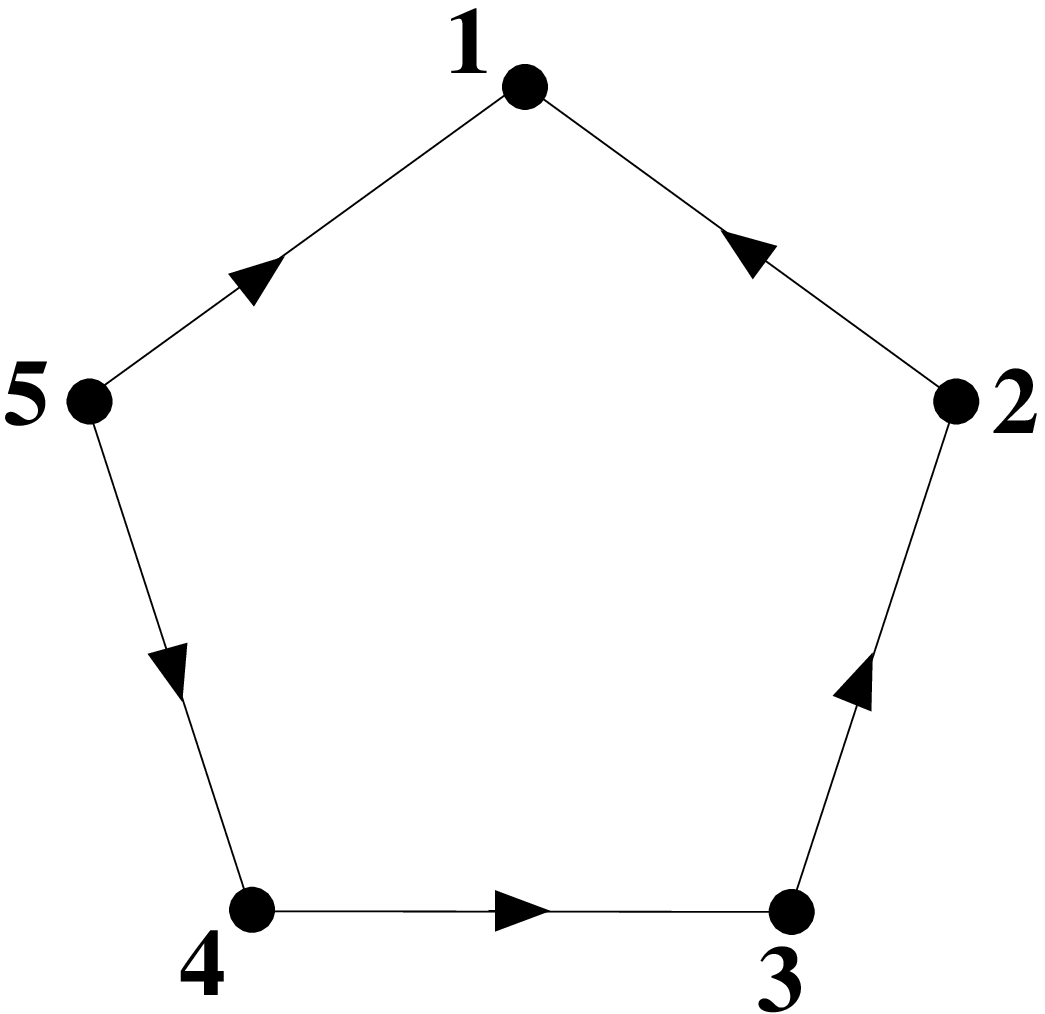}\label{subfig:P51}
}\qquad \subfigure[$P_5^{(2)}$]{
\includegraphics[width=3cm]{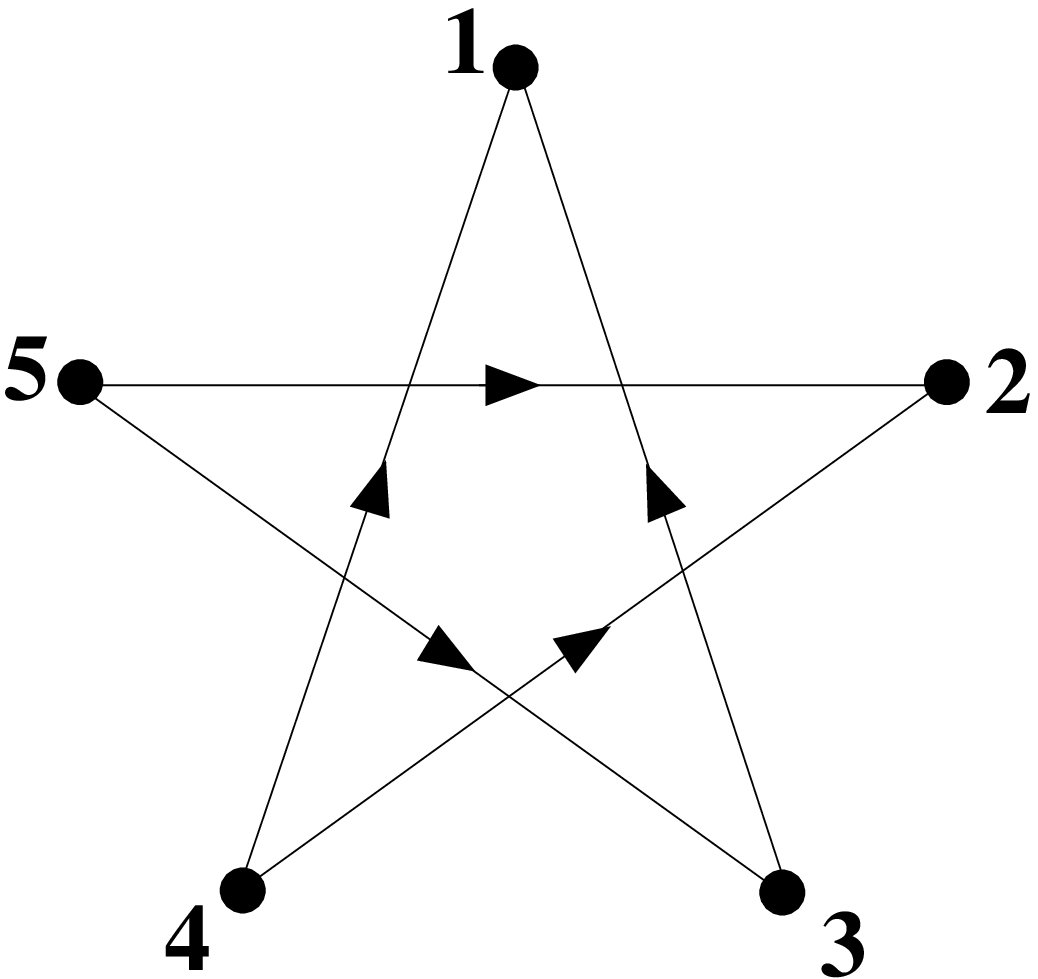}\label{subfig:P52}
} \caption{The period $1$ primitives for $4$ and $5$ nodes.} \label{45node}
\end{figure}

Figure \ref{45node} shows these primitives for $N=4,5$.

\subsubsection{The General Period $1$ Quiver}

We start with a skew-symmetric matrix whose first column is of the form $(0,m_1,\dots ,m_{N-1})^T$ and {\em construct} the remaining elements $b_{ij}$.  For $1\leq i,j\leq N-1$, let
$$
\varepsilon_{ij}=\frac{1}{2}(m_i|m_j|-m_j|m_i|).
$$
Then if $m_i$ and $m_j$ have the same sign, $\varepsilon_{ij}=0$. Otherwise
$\varepsilon_{ij}=\pm |m_im_j|$, where the sign is that of $m_i$.  The general formula (\ref{gen-mut}) for matrix mutation then gives
$$
\widetilde{B}=\mu_1 B \quad\Rightarrow\quad \tilde{b}_{ij}=b_{ij}+\varepsilon_{i-1,j-1}.
$$
We also have that the effect of the rotation $B\mapsto \rho B \rho^{-1}$ is to move the
entries of $B$ down and right one step, so that $(\rho B \rho^{-1})_{ij}=b_{i-1,j-1}$, remembering that indices are
labelled modulo $N$, so $N+1\equiv 1$.

Therefore, $\mu_1 B=\rho B\rho^{-1}$ implies
$$
 b_{ij} = b_{i-1,j-1}+\varepsilon_{j-1,i-1}  \quad\mbox{and}\quad (b_{N,1},b_{N,2},\ldots ,b_{N,N-1})=(m_1,\dots ,m_{N-1}).
$$
The first of these is a difference equation, building the lower triangular elements of $B$, with initial conditions on the left column. The second of these is a boundary condition.  Together, they imply that $m_{n-r}=m_r$ (the {\em palindromic property}) and that
\be
b_{ij} = b_{i-j+1,1}+\varepsilon_{j-1,i-1}+\varepsilon_{j-2,i-2}+\cdots +\varepsilon_{1,i-j+1}.
\ee

\paragraph{A Description in Terms of Primitives.}

Recall that for an even (or odd) number of nodes, $N=2r$ (or
$N=2r+1$), there are $r$ {\em primitives}, labelled $B_{2r}^{(k)}$ (or
$B_{2r+1}^{(k)}$), $k=1,\cdots ,r$.
We denote the general linear combination of these by
$$
\widetilde{B}_{2r}(\mu_1,\dots ,\mu_r) = \sum_{j=1}^r \mu_j B_{2r}^{(j)},
\quad\mbox{or}\quad
\widetilde{B}_{2r+1}(\mu_1,\dots ,\mu_r) = \sum_{j=1}^r \mu_j
B_{2r+1}^{(j)},
$$
for integers $\mu_j$.  The quivers corresponding to
$\widetilde{B}_{2r}(\mu_1,\dots ,\mu_r)$ and
$\widetilde{B}_{2r+1}(\mu_1,\dots ,\mu_r)$
do not have periodicity properties (unless all $\mu_j$ have the same sign).

\bt[The general period $1$ quiver]  \label{p1-theorem}
Let $B_{2r}$ (respectively $B_{2r+1}$) denote the matrix corresponding
to the general even (respectively odd) node quiver of mutation periodicity
$1$.  Then
\begin{enumerate}
\item
$$
B_{2r} = \widetilde{B}_{2r}(m_1,\dots ,m_r) +
  \sum_{k=1}^{r-1} \widetilde{B}_{2(r-k)}(\varepsilon_{k,k+1},\dots ,\varepsilon_{kr}) ,
$$
where
$\widetilde{B}_{2(r-k)}(\varepsilon_{k,k+1},\dots ,\varepsilon_{kr})$ is embedded
in a $2r\times 2r$ matrix in rows and columns $k+1,\dots ,2r-k$.    %
\item
$$
B_{2r+1} = \widetilde{B}_{2r+1}(m_1,\dots ,m_r) +
  \sum_{k=1}^{r -1}
\widetilde{B}_{2(r-k)+1}(\varepsilon_{k,k+1},\dots ,\varepsilon_{kr}) ,
$$
where
$\widetilde{B}_{2(r-k)+1}(\varepsilon_{k,k+1},\dots ,\varepsilon_{kr})$ is
embedded in a $(2r+1)\times (2r+1)$ matrix in
rows and columns $k+1,\dots ,2r+1-k$.    %
\end{enumerate}
\et  %

\br  %
From these formulae, together with (\ref{bnk}), we see that the {\em first column} of $B_N$, corresponding to
a period $1$ quiver, is palindromic, with $b_{r1}=m_{r-1}$ and $m_{N-r}=m_r$.  These are the only matrix elements
which enter the formula of the cluster exchange relation, and thus the recurrence formula.  The matrix $B$ has the following
symmetry: $b_{N-j+1,N-i+1}=b_{ij}$.
\er  %

\subsubsection{Recurrences from Period $1$ Solutions}\label{p1-recur}

Corresponding to each period $1$ quiver
\be\label{arec}
x_{n+N}\, x_n = \prod_{m_j\geq 0}x_{n+j}^{m_j}+ \prod_{m_j\leq 0}x_{n+j}^{-m_j},
\ee
where the indices in each product lie in the range $1\leq j\leq N-1$, with the
exponents $(m_1,...,m_{N-1})$ forming an integer $(N-1)$-tuple which is
palindromic, so that $m_j = m_{N-j}$.

\bex[The Somos-$4$ Quiver] {\em  %
For the Somos-$4$ quiver, we have $m_1=1, m_2=-2$
and our formula requires the addition of a further two arrows between nodes $3$
and $2$ (see Figure \ref{s4sum}).
\begin{figure}[htb]
\centering \subfigure[$P_4^{(1)}$]{
\includegraphics[width=2cm]{P41.eps}
}\qquad \subfigure[$P_4^{(2)}$]{
\includegraphics[width=2cm]{P42.eps}
}\qquad \subfigure[$P_2^{(1)}$]{
\includegraphics[width=2cm]{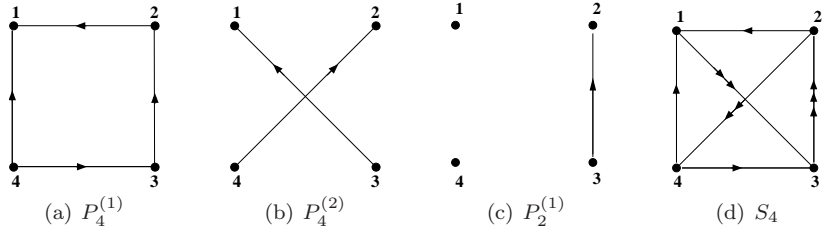}
}\qquad \subfigure[$S_4$]{
\includegraphics[width=2cm]{somos4quiver.eps}
}\caption{One of $P_4^{(1)}$ minus two of $P_4^{(2)}$ plus two of $P_2^{(1)}$
gives $S_4$}\label{s4sum}
\end{figure}
}\eex  %

\bex[The Somos-$5$ Quiver] {\em  %
For the Somos-$5$ quiver, we have $m_1=1, m_2=-1$, giving
\be\label{somos5}  %
x_n x_{n+5}=x_{n+1}x_{n+4}+x_{n+2}x_{n+3}.
\ee  %
The full quiver is given by $P_5^{(1)}-P_5^{(2)}+P_3^{(1)}$
(see Figure \ref{fig:somos5quiver}).
\begin{figure}[ht]
\centering
\includegraphics[width=4cm]{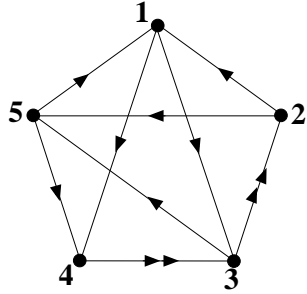}
\caption{The Somos-$5$ quiver.}\label{fig:somos5quiver}
\end{figure}
}\eex   %

\bex[Period $1$ Quiver with $6$ Nodes]  \label{n=6p1}  {\em %
Here the matrix has the form
\bea   %
B &=& \left(\begin{array}{cccccc}
     0 & -m_1 & -m_2 & -m_3 & -m_2 & -m_1 \\
     m_1 & 0 & -m_1 & -m_2 & -m_3 & -m_2 \\
     m_2 & m_1 & 0 & -m_1 & -m_2 & -m_3 \\
     m_3 & m_2 & m_1 & 0 & -m_1 & -m_2 \\
     m_2 & m_3 & m_2 & m_1 & 0 & -m_1 \\
     m_1 & m_2 & m_3 & m_2 & m_1 & 0
     \end{array}\right)   \nn\\[3mm]
     && \quad +
     \left(\begin{array}{c|cccc|c}
     0 & 0 & 0 & 0 & 0 & 0 \\
     \hline
     0 & 0 & -\varepsilon_{12} & -\varepsilon_{13} & -\varepsilon_{12} & 0 \\
     0 & \varepsilon_{12} & 0 & -\varepsilon_{12} & -\varepsilon_{13} & 0 \\
     0 & \varepsilon_{13} & \varepsilon_{12} & 0 & -\varepsilon_{12} & 0 \\
     0 & \varepsilon_{12} & \varepsilon_{13} & \varepsilon_{12} & 0 & 0 \\
     \hline
     0 & 0 & 0 & 0 & 0 & 0
     \end{array}\right)  +
     \left(\begin{array}{cc|cc|cc}
     0 & 0 & 0 & 0 & 0 & 0 \\
     0 & 0 & 0 & 0 & 0 & 0 \\
     \hline
     0 & 0 & 0 & -\varepsilon_{23} & 0 & 0 \\
     0 & 0 & \varepsilon_{23} & 0 & 0 & 0 \\
     \hline
     0 & 0 & 0 & 0 & 0 & 0 \\
     0 & 0 & 0 & 0 & 0 & 0
     \end{array}\right)   , \nn
\eea   %
which can be written as
$$
B = \sum_{j=1}^3 m_k\, B_6^{(k)} + \sum_{k=1}^2 \varepsilon_{1,k+1}\, B_4^{(k)} +
   \varepsilon_{23}\, B_2^{(1)} ,
$$
where the periodic solutions with fewer rows and columns are embedded
symmetrically within a $6\times 6$ matrix.

Some specific examples of this can be found in the next example.
}\eex   %

\bex[Gale-Robinson Sequence ($N$ nodes)]\label{galerob}  {\em   %
The $2-$term Gale-Robinson recurrence is given by
$$
x_nx_{n+N}=x_{n+N-r}x_{n+r}+x_{n+N-s}x_{n+s} ,
$$
for $0<r<s\leq N/2$, and is one of the examples discussed in \cite{02-2}.
This corresponds to the period $1$ quiver with $m_r=1$ and $m_s=-1$
(unless $N=2s$, in which case we take $m_s=-2$).

\paragraph{Subcases of Somos-$6$ when $N=6$:}
If we choose $m_1=1,\, m_2=-1,\, m_3=0$, the full quiver is $P_6^{(1)}-P_6^{(2)}+P_4^{(1)}$
(see Figure \ref{subfig:gr1}) and corresponds to the recurrence
\be\label{gr1}  %
x_n x_{n+6}=x_{n+1}x_{n+5}+x_{n+2}x_{n+4}.
\ee  %
If we choose $m_1=1,\, m_2=0,\, m_3=-2$, the full quiver is $P_6^{(1)}-2P_6^{(3)}+2P_4^{(2)}$
(see Figure \ref{subfig:gr2}) and corresponds to the recurrence
\be\label{gr2}  %
x_n x_{n+6}=x_{n+1}x_{n+5}+x_{n+3}^2.
\ee  %
If we choose $m_1=0,\, m_2=1,\, m_3=-2$, the full quiver is $P_6^{(2)}-2P_6^{(3)}+2P_2^{(1)}$
(see Figure \ref{subfig:gr3}) and corresponds to the recurrence
\be\label{gr3}  %
x_n x_{n+6}=x_{n+2}x_{n+4}+x_{n+3}^2.
\ee  %
\begin{figure}[htb]
\centering
\psfrag{1}{$1$}\psfrag{2}{$2$}\psfrag{3}{$3$}\psfrag{4}{$4$}\psfrag{5}{$5$}\psfrag{6}{$6$}
\subfigure[Equation (\ref{gr1})]{
\includegraphics[width=4cm]{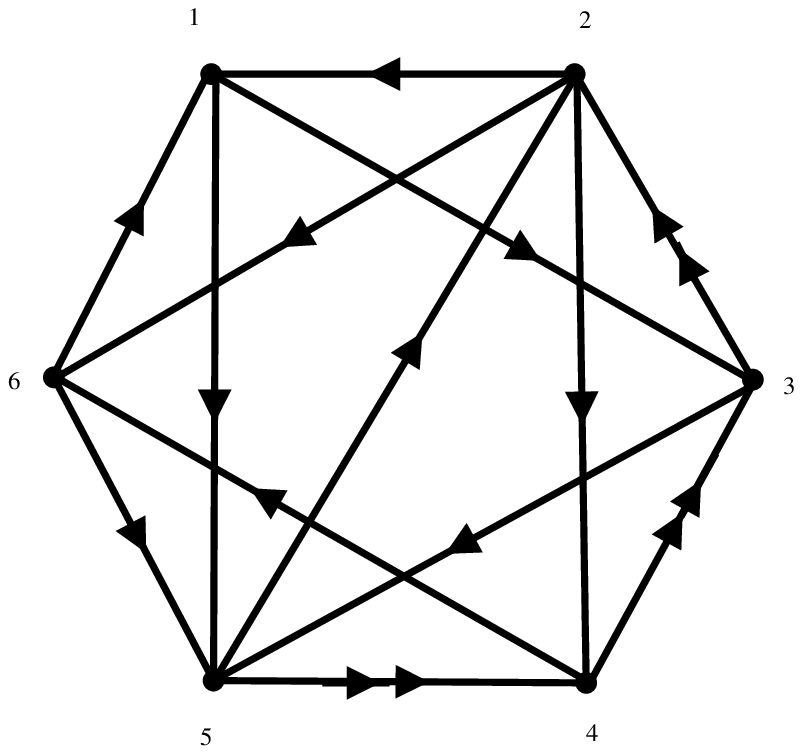}\label{subfig:gr1}
}\qquad \subfigure[Equation (\ref{gr2})]{
\includegraphics[width=4cm]{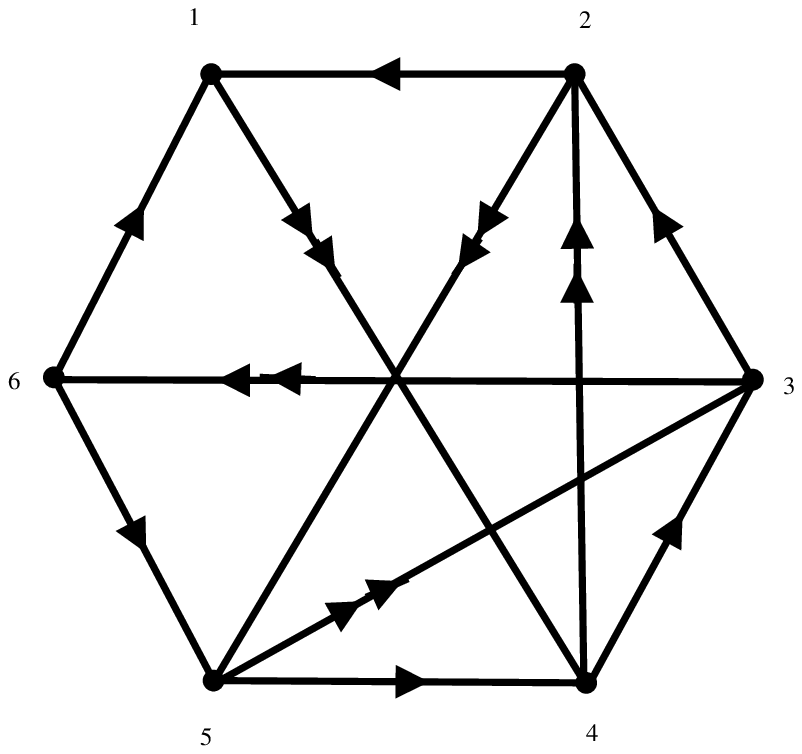}\label{subfig:gr2}
}\qquad \subfigure[Equation (\ref{gr2})]{
\includegraphics[width=4cm]{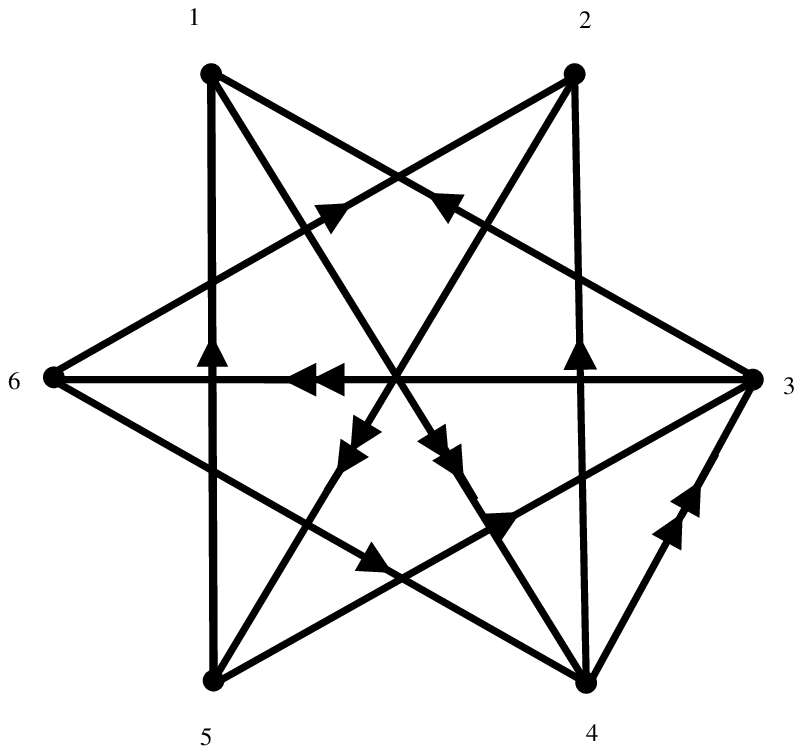}\label{subfig:gr3}
}\caption{Quivers giving subcases of Somos-$6$ }\label{Fig:GaleRob}
\end{figure}
}\eex  %

\subsection{Period $2$ Quivers}

Already at period $2$, we cannot give a full classification of the possible
quivers. However, we can give the {\em full list} for low values of $N$, the
number of nodes.  We can also give a class of period $2$ quivers which exists
for all $N$.

Whereas period $1$ quivers gave rise to recurrences in a single variable $x_n$, period
$2$ quivers give us two components $(x_n,y_n)$.

\subsubsection{A Family of Period $2$ Solutions}\label{p2-reg}

It is possible to modify the derivation of the general period $1$ quiver, to incorporate
an involution so as to give a corresponding family of period $2$ quivers.

We start with a skew-symmetric matrix whose first column is of the form $(0,m_1,\dots ,m_{N-1})^T$
and {\em impose} the condition $m_{N-r}=m_r, \, r\neq 1$ and $m_{N-1}=m_{\overline{1}}\neq m_1$.  We
require $\varepsilon_{1\overline{1}}=0$, so assume both $m_1\ge 0$ and $m_{\overline{1}}\ge 0$.
We write $B=B(\mathbf{m})=B(m_1,m_2,\ldots ,m_{N-2}, m_{\overline{1}})$ and define
$\sigma(\mathbf{m})=\sigma(m_1,m_2,\ldots ,m_{N-2}, m_{\overline{1}})
              =(m_{\overline{1}},m_2,\ldots ,m_{N-2},m_1)$.

We construct a matrix $B=B(\mathbf{m})$ which satisfies the equation
\begin{equation}
\mu_1(B)=\rho B(\sigma(\mathbf{m}))\rho^{-1}.\label{eqn:sigmad}
\end{equation}
Since $\sigma$ is an involution, we obtain period $2$ solutions in this way.
This condition leads to the formula
\begin{equation} \label{eqn:period2equation}
b_{ij}=\sigma(b_{i-1,j-1})+\varepsilon_{j-1,i-1},
\end{equation}
with which we iteratively build the matrix $B$ from the left column.  Condition (\ref{eqn:sigmad}) also
implies that $(b_{N1},b_{N2},\ldots ,b_{N,N-1})=\sigma(\mathbf{m})$, which acts as a boundary condition for the difference
equation (\ref{eqn:period2equation}).  This puts constraints on the possible values of $m_i$.  We require $m_r\geq 0$ for $r$ odd, $r\geq 3$
and, in addition, $m_2=-1$ for $N$ odd.  To achieve nontrivial mutation, we must choose at least one of the $m_r<0$, for $r$ even.

\bex[The $4$ Node Case]\label{reg4}  {\em   %
We take $m_1=r>0,\, m_2=-s<0,\, m_3=m_{\overline{1}}=t>0$, with $t\neq r$ and easily construct
$$
B(1)=\left(
       \begin{array}{cccc}
         0 & -r & s & -t \\
         r & 0 & -t-rs & s \\
         -s & t+rs & 0 & -r \\
         t & -s & r & 0 \\
       \end{array}
     \right), \quad
     B(2)=\left(
       \begin{array}{cccc}
         0 & r & -s & t \\
         -r & 0 & -t & s \\
         s & t & 0 & -r-st \\
         -t & -s & r+st & 0 \\
       \end{array}
     \right),
$$
satisfying (\ref{eqn:sigmad}).
}\eex

\bex[The $5$ Node Case] \label{reg5} {\em   %
We take $m_1=r>0,\, m_2=-1,\, m_3=m_{\overline{1}}=t>0$, with $t\neq r$ and easily construct
\bea   %
B(1) &=& \left(\begin{array}{ccccc}
     0 & -r & 1 & 1 & -t \\
     r & 0 & -r-t & 1-r & 1 \\
     -1 & r+t & 0 & -r-t & 1 \\
     -1 & r-1  & r+t & 0 & -r \\
     t & -1 & -1 & r & 0
     \end{array}\right) , \nn\\
     && \label{n5-1pos4pos}   \\
B(2) &=& \left(\begin{array}{ccccc}
     0 & r & -1 & -1 & t \\
     -r & 0 & -t & 1 & 1 \\
     1 & t & 0 & -r-t & 1-t \\
     1 & -1  & r+t & 0 & -r-t \\
     -t & -1 & t-1 & r+t & 0
     \end{array}\right) ,   \nn
\eea   %
satisfying (\ref{eqn:sigmad}).
}\eex

\subsubsection{``Exceptional'' Period $2$ Solutions}\label{p2-excep}

For low values of $N$ it is possible to determine all period $2$ quivers.  For $N\geq 5$ there exist quivers that are {\em not} in the ``regular'' class, described in section \ref{p2-reg}.  We call these ``exceptional solutions''.  I give one example here.

\bex[An Exceptional, Period $2$, $5$ Node Quiver]  {\em   %
Here we have
\bea   %
B(1) &=& \left(\begin{array}{ccccc}
     0 & -m_1 & -1 & -m_1-1 & 1 \\
     m_1 & 0 & 1 & -m_1-1 & -m_1-1 \\
     1 & -1 & 0 & 1 & -1 \\
     m_1+1 & m_1+1  & -1 & 0 & -m_1 \\
     -1 & m_1+1 & 1 & m_1 & 0
     \end{array}\right) , \nn\\
     && \label{n5-1pos4neg2pos}   \\
B(2) &=& \left(\begin{array}{ccccc}
     0 & m_1 & 1 & m_1+1 & -1 \\
     -m_1 & 0 & 1 & -m_1-1 & -1 \\
     -1 & -1 & 0 & 1 & 0 \\
     -m_1-1 & m_1+1  & -1 & 0 & 1 \\
      1 & 1 & 0 & -1 & 0
     \end{array}\right) .   \nn
\eea   %
}\eex

\subsubsection{Recurrences from Period $2$ Solutions}\label{p2-recur}

Consider (\ref{periodchain}), with $m=2$.  The configuration of arrows at node $2$ of $Q(2)$ will be different from that at node $1$ of $Q(1)$, so the cluster exchange relation will be different.  However, node $3$ of $Q(3)=\rho^2 Q(1)$ will have the same configuration of arrows as node $1$ of $Q(1)$.  Hence, after two mutations we obtain the {\em same} formula, but with the index shifted.  The result is that we have a pair of formulae which alternate as we go around the nodes.  It is thus natural to relabel the cluster variables accordingly.

Labelling the odd and even nodes by $x$ and $y$, we start with an initial cluster
$(x_1,y_1,x_2,y_2,\dots)$ at nodes $1, 2, 3, 4, \dots$.  We see immediately that there is a ``mismatch'' if $N$ is odd.  This has a consequence for the initial value problem.  When $N=2m$, the recurrence is of order $m$.  When $N=2m-1$, the recurrence is again of order $m$, but the first exchange relation plays the role of a {\bf boundary condition}.

\bex[The $4$ Node Case]\label{reg4rec}  {\em   %
For Example \ref{reg4}, we combine the two formulae to give a recurrence
$(x_n,y_n)\mapsto (x_{n+2},y_{n+2})$, given by
\be\label{period2-4xy}  %
x_n x_{n+2}=y_n^ry_{n+1}^t+x_{n+1}^s,\quad
y_n y_{n+2} = x_{n+2}^r x_{n+1}^t+y_{n+1}^s = \frac{(y_n^ry_{n+1}^t+x_{n+1}^s)^r x_{n+1}^t}{x_n^r}+y_{n+1}^s.
\ee  %
}\eex  %

\begin{figure}[hbt]
\begin{center}
\unitlength=0.5mm
\begin{picture}(240,55)
%==========nodes==============
\put(10,45){\makebox(0,0){$\rho^{-1}\, Q(2)$}}
\put(70,45){\makebox(0,0){$\rho\,
Q(1)$}} \put(130,45){\makebox(0,0){$\rho\, Q(2)$}}
\put(10,5){\makebox(0,0){$Q(1)$}} \put(70,5){\makebox(0,0){$Q(2)$}}
\put(130,5){\makebox(0,0){$\rho^2\, Q(1)$}} \put(190,5){\makebox(0,0){$\rho^2\,
Q(2)$}}
%=========vectors========================
\put(25,45){\vector(1,0){30}} \put(80,45){\vector(1,0){35}}
\put(140,45){\vector(1,0){35}} \put(20,5){\vector(1,0){35}}
\put(80,5){\vector(1,0){35}} \put(140,5){\vector(1,0){35}}
\put(200,5){\vector(1,0){35}} \put(60,10){\vector(-3,2){40}}
\put(120,10){\vector(-3,2){40}} \put(180,10){\vector(-3,2){40}}
%==========labels==============================
\put(40,50){\makebox(0,0){$\mu_1$}} \put(95,50){\makebox(0,0){$\mu_2$}}
\put(160,50){\makebox(0,0){$\mu_3$}} \put(40,0){\makebox(0,0){$\mu_1$}}
\put(95,0){\makebox(0,0){$\mu_2$}} \put(160,0){\makebox(0,0){$\mu_3$}}
\put(220,0){\makebox(0,0){$\mu_4$}} \put(45,30){\makebox(0,0){$\rho^{-1}$}}
\put(105,30){\makebox(0,0){$\rho^{-1}$}}
\put(165,30){\makebox(0,0){$\rho^{-1}$}}
\end{picture}
\end{center}
\caption{Period $2$ quivers and mutations} \label{p2quivers}
\end{figure}
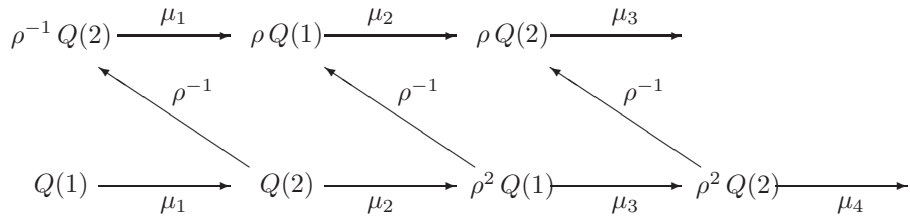

\br  %
We can construct a different recurrence by first rotating $B(2)$, with $B(2)\mapsto \rho^{-1} B(2)\rho$ (so node $2$ becomes
node $1$).  Mutation at this new node $1$ then gives a rotated version of the old quiver $Q(1)$, so we just cycle through
the same pair of quivers (see Figure \ref{p2quivers}).
The rotation $\rho^{-1}\, Q(2)$ is just $Q(1)$ with $r\leftrightarrow t$, so the second form of the recurrence is just (\ref{period2-4xy}), with $r\leftrightarrow t$.
\er  %

\bex[A $5$ Node Case]  \label{5nodep2}  {\em   %
Consider the case with matrices (\ref{n5-1pos4pos}).  The same procedure leads first to the relation
$$
x_1 y_3=y_1^r x_3^t + x_2 y_2,
$$
after which the variables satisfy a third order ($2-$component) recurrence
\be  \label{xnyn5node}  %
 y_n x_{n+3} = y_{n+2}^r x_{n+1}^t+x_{n+2} y_{n+1},  \quad
   x_{n+1} y_{n+3} = y_{n+1}^r x_{n+3}^t + x_{n+2} y_{n+2}, \;\;\; n=1,2,\cdots .
\ee   %
We start with initial conditions $(x_1,y_1,x_2,y_2,x_3)=(c_1,c_2,c_3,c_4,c_5)$.  The first relation gives $y_3$, which is the {\em sixth} initial condition required for the recurrence (\ref{xnyn5node}).

As above, it is possible to construct a companion recurrence, corresponding to the choice
$$
\bar B(1) = \rho^{-1} B(2) \rho , \quad  \bar B(2) = \rho B(1) \rho^{-1}.
$$
}\eex   %

\section{Symmetries and Complete Integrability}\label{poisson}
\setcounter{equation}{0}

The main purpose of this section is to arrive at a general definition of complete integrability for a discrete dynamical system defined by a map.
This definition requires the notion of a {\em continuous symmetry}, so we first discuss the Lie Bracket and commutativity of vector fields.
The culmination of this part of the discussion is {\em Lie's Theorem}, which states that in $n$ dimensions, $n$ commuting vector fields form a completely integrable system, and can be {\em simultaneously straightened}.

When applied to a Hamiltonian system, this leads to Liouville's Theorem of complete integrability.  For this part of the discussion we first give the general definition of a Poisson bracket and Poisson manifold.  Poisson commutativity of two functions gives the commutativity of the corresponding Hamiltonian vector fields.  For complete integrability of a map we require some additional invariance properties.

In the case of cluster maps, the matrix $B$ defines a pre-symplectic structure, which is invariant under the action of the map when $B$ is mutation periodic.  This then enables the construction of an invariant Poisson bracket, possibly after the reduction to a lower dimensional space.

\subsection{The Lie Bracket and Commuting Vector Fields}

Let $M$ be an $n-$dimensional manifold, and suppose $(x_1,\dots ,x_n)$ is a local coordinate system at point $p\in M$.
Any collection of smooth functions $\{X_i({\bf x})\}_{i=1}^n$ in some neighbourhood of $M$ defines a smooth vector field, which we can associate with a first order partial differential operator (acting on smooth functions $f:M\rightarrow \mathbb{R}$)
\be\label{Xf}  %
{\bf X} f = \sum_{i=1}^n X_i \frac{\pa f}{\pa x_i} .
\ee  %
Integral curves are any trajectories $x_i(t)$, satisfying $\frac{dx_i}{dt}=X_i$.

\bd[The Commutator]  %
 The commutator $[{\bf X},{\bf Y} ]={\bf X}\circ {\bf Y}-{\bf Y}\circ {\bf X}$
 of two vector fields, ${\bf X} = \sum_{i=1}^n X_i \frac{\pa}{\pa x_i}$ and
 ${\bf Y} = \sum_{i=1}^n Y_i \frac{\pa}{\pa x_i}$,
is the vector field with components:
\be \label{liebra} %
[ {\bf X} , {\bf Y} ]_j = \sum_{i=1}^n \left(
         X_i \frac{\pa Y_j}{\pa x_i} -  Y_i \frac{\pa X_j}{\pa x_i}
                         \right) =  {\bf X}(Y_j) - {\bf Y}(X_j) .
\ee   %
\ed  %
The commutator is also called the {\em Lie bracket}. It is {\em bilinear}, {\em skew-symmetric} and satisfies the {\em Jacobi identities}, so defines a {\em Lie algebraic} structure on vector fields.

Clearly, the coordinate vector fields ${\bf e}_i=\frac{\pa}{\pa x_i}$ commute: $[{\bf e}_i,{\bf e}_j]=0$.  This means that their respective integral curves can be used as coordinate curves, since, from a given initial point, we unambiguously arrive at the same point, by following first one and then the other, regardless of the order we choose.  Furthermore, if two vector fields commute ($[{\bf X},{\bf Y}]=0$), then this property still holds, which means that we can {\em simultaneously straighten} the vector fields.  This means that there exist coordinates $y_i$, such that ${\bf X}=\frac{\pa}{\pa y_1}$ and ${\bf Y}=\frac{\pa}{\pa y_2}$.  Generally, we have:

\bt[Complete Integrability (Lie's Theorem)] \label{lie}  %
Let ${\bf X}_1,\dots , {\bf X}_n$ be $n$ {\em independent} vector fields on
$\mathbb{R}^n$:
$$
{\bf X}_i = \sum_{j=1}^n a_{ij}\, \frac{\pa}{\pa x_j} .
$$
If $[{\bf X}_i,{\bf X}_j]=0$, for all $i,j$, then we can {\bf simultaneously
straighten} these $n$ vector fields.
\et    %
\begin{prf}
Our task is to solve the linear equations
\be                   \label{ay=del}
{\bf X}_i y_j = \delta_{ij}\quad\Rightarrow\quad \sum_{k=1}^n a_{ik} \frac{\pa y_j}{\pa x_k} = \delta_{ij} ,
\ee
which reduce to
\be                        \label{dydxi}   %
\frac{\pa y_j}{\pa x_i} = b_{ij}, \quad\mbox{where}\quad B = A^{-1}.
\ee   %
The integrability conditions:
$$
\frac{\pa b_{ij}}{\pa x_k} = \frac{\pa b_{kj}}{\pa x_i}
$$
are guaranteed by the commutativity of the vector fields.  Since $b_{ij}$ are just functions of $\bf x$, this reduces to quadratures, which can be explicitly integrated for simple vector fields.
\end{prf}

\subsection{Poisson Manifolds}

We now suppose there exists an algebraic operation $\{f,g\}$ between functions on $M$, possessing the following properties, thus defining a Poisson bracket on $M$.

\bd[A Poisson Bracket]\label{pbdef} Let $f,\,g$ and $h$ be functions, $\alpha$ and $\beta$ constants.
Then the operation $\{f,g\}$ is said to be a Poisson bracket if the following properties hold:
\begin{tabbing}
{\bf 1. Bilinearity:}  \qquad\qquad \= $\{ \alpha f + \beta g, h \} = \alpha\,\{ f,h \} + \beta\,\{ g,h \}$,  \\[1mm]
{\bf 2. Skew Symmetry:}   \> $\{f,g\} = - \{g,f\}$,  \\[1mm]
{\bf 3. Jacobi Identity}\>     $\{\{f,g\},h\}+\{\{g,h\},f\}+\{\{h,f\},g\}=0$,\\[1mm]
{\bf 4. Leibnitz Rule}  \> $ \{fg,h\} = \{f,h\}g + f\{g,h\}$.
\end{tabbing}
\ed   %
The first three define a Lie algebra on the functions on $M$.  The fourth guarantees the existence of a {\em Hamiltonian vector field}:
$$
{\bf X}_h f = \{f,h\}\quad\Rightarrow\quad  {\bf X}_h (fg)= ({\bf X}_h f) g + f ({\bf X}_h g).
$$

\bd[Poisson Tensor]  \label{def:Ptensor}  %
The Poisson tensor $P({\bf x})$, is defined by its matrix of coefficients:
$$
P_{ij} = \{x_i,x_j\},
$$
which is skew-symmetric: $P_{ji}=-P_{ij}$.
\ed  %
\br  %
The Poisson tensor is a bivector so in standard tensor notation would have \underline{upper} indices (as should $x_k$), but we have avoided this notation here.  This accounts for the formula for the Lie derivative, given below.
\er  %

The Jacobi identity implies:
$$
\{\{x_i,x_j\},x_k\}+\{\{x_j,x_k\},x_i\}+\{\{x_k,x_i\},x_j\}=0 ,\quad\mbox{for}\;\; i<j<k,
$$
which then imply
\be\label{pdp}  %
\sum_{\ell=1}^n \left(P_{i\ell}\frac{\pa P_{jk}}{\pa x_\ell} +P_{j\ell}\frac{\pa P_{ki}}{\pa x_\ell} +
                                 P_{k\ell}\frac{\pa P_{ij}}{\pa x_\ell}\right)=0 ,\quad\mbox{for}\;\; i<j<k.
\ee  %
These conditions are implied by the properties given in the definition of an abstract Poisson bracket.  In applications, we \underline{define} a (specific) Poisson bracket by presenting a skew-symmetric matrix $P_{ij}$, satisfying conditions (\ref{pdp}).

The $i^{th}$ component of ${\bf X}_h$ is just ${\bf X}_h x_i = \{x_i,h\}=P_{ij}\frac{\pa h}{\pa x_j}$, so the {\em Hamiltonian vector field} and {\em Poisson bracket} formulae are given by
\be\label{fph} %
{\bf X}_h= \sum_{i,j=1}^n P_{ij}\frac{\pa h}{\pa x_j} \frac{\pa }{\pa x_i},\quad\mbox{and}\quad
\{f,h\}= \sum_{i,j} \frac{\pa f}{\pa x_i}P_{ij}\frac{\pa h}{\pa x_j}.
\ee  %

\bd[Casimirs and Rank of a Poisson Tensor]
Suppose the rank of the Poisson tensor $P$ is $2r$, so $n=2r+m$.  When $m=0$, the Poisson bracket is non-degenerate and its inverse defines a symplectic form.  If $m>0$ and the $m$ null vectors can be written as gradients of some functions ${\cal C}_k, \, k=1,\dots m$, then these functions are said to be Casimirs. The Casimir functions satisfy
$$
\{f,{\cal C}_k\}=0 \quad\mbox{for arbitrary function}\;\; f.
$$
\ed

\paragraph{The Lie Derivative of a Poisson Tensor along $\bf X$} is defined by the formula
$$
(L_{X} P)_{ij} = \sum_k \left(X_k\frac{\pa P_{ij}}{\pa x_k}-P_{ik}\frac{\pa X_j}{\pa x_k}-P_{kj}\frac{\pa X_i}{\pa x_k}\right).
$$
In the particular case of $X_h$, we have
$$
(L_{X_h} P)_{ij} = -\sum_{k,\ell} \left(P_{\ell k}\frac{\pa P_{ij}}{\pa x_k}+P_{ik}\frac{\pa P_{j\ell}}{\pa x_k}+P_{kj}\frac{\pa P_{\ell i}}{\pa x_k}\right)\frac{\pa h}{\pa x_\ell}.
$$
We can interpret this formula in two ways
\begin{enumerate}
\item  If $P$ is a Poisson tensor, then for any Hamiltonian vector field $X_h$, $L_{X_h} P=0$.
\item  If {\em for any} function $h$, $L_{X_h} P=0$, where $X_h$ is defined by formula (\ref{fph}), for some skew-symmetric (contravariant) tensor $P$, then $P$ satisfies the Jacobi identities, so defines a Poisson bracket.
\end{enumerate}

\subsubsection{The Commutator of Two Hamiltonian Vector Fields}

To each function $f({\bf x})$ there corresponds a Hamiltonian vector field (\ref{fph}).
The {\em commutator} of two such vector fields is defined by (\ref{liebra}).  The Jacobi identities give us an important relation between two Lie algebras.
\bt   %
If $f$ and $g$ are any $2$ functions on phase space, then:
\be     \label{jacob}   %
[ X_f, X_g ] = -X_{\{f,g\}} .
\ee   %
\et    %
We thus have that the commutator of $2$ {Hamiltonian} vector fields is not just
a vector field, but also Hamiltonian.  The formula (\ref{jacob}) is perhaps the
most important consequence of the Jacobi identities, giving us that
the flows generated by Poisson commuting functions commute:
\be\label{xfxg=0}  %
\{f,g\}=0\quad\Rightarrow\quad [X_f,X_g]=0.
\ee  %

\subsubsection{Complete Integrability of a Hamiltonian System}

We now describe Liouville integrability, which is a corollary to Lie's Theorem \ref{lie}. See \cite{78-3,04-4} for more details.

We start with a Poisson tensor $P$ of rank $2r$ ($n=2r+m$), with $m$ Casimir functions $\{{\cal C}_k\}_{k=1}^m$, satisfying $\{{\cal C}_k,f\}=0$, for all functions $f({\bf x})$.  This last statement means that all Hamiltonian flows generated by $P$ are tangent to the level surfaces of ${\cal C}_k$, of dimension $2r$:
$$
M_s=\{{\bf x}\in M: {\cal C}_k=s_k\},\quad s_k \;\;\mbox{constants},
$$
determined by the initial conditions.  The matrix $\frac{\pa {\cal C}_j}{\pa x_i},\, i=1,\dots ,n,\, j=1,\dots , m$ will have maximum rank $m$ {\em almost everywhere}.  This condition will fail at most on some lower dimensional surfaces within $M$.  The regular open regions of maximal rank are foliated by the surfaces $M_s$, called {\em symplectic leaves}.  If we use ${\cal C}_i$ as local coordinates in such a region, then the matrix $P$ will have a $2r\times 2r$ non-degenerate block, with the remainder of the matrix being an array of zeros.

Now consider such a symplectic leaf, with local coordinates $x_1, \dots , x_{2r}$.  Suppose we now have $r$ {\em independent} functions $h_i, i=1,\dots , r$ (again the matrix $\frac{\pa h_j}{\pa x_i},\, i=1,\dots ,2r,\, j=1,\dots , r$ must be of maximal rank $r$), which are {\em in involution}:
\be\label{involutive}  %
\{h_i,h_j\}=0, \quad\mbox{for all}\;\; i,j = 1,\dots , r.
\ee  %
If this is written
$$
X_{h_j} h_i = 0,\quad \mbox{for}\;\; i=1,\dots , r,
$$
we see that, for each $j=1,\dots , r$, $X_{h_j}$ is tangent to the level surface
$$
M_c=\{{\bf x}\in M_s: h_i({\bf x})=c_i\},\quad c_i \;\;\mbox{constants},
$$
determined by initial conditions.
Furthermore, by (\ref{jacob}), the $r$ vector fields $X_{h_j}$ {\em commute} on this $r$ dimensional surface $M_c$.
By Lie's Theorem \ref{lie} these vector fields form a completely integrable system, so can be ``simultaneously straightened'', forming coordinate curves within $M_c$.

\bt[Liouville Integrability] \label{liouville} %
Let $h_1,\dots ,h_r$ be $r$ independent functions, satisfying (\ref{involutive}) on the $2r$ dimensional manifold $M_s$.  Then each of the Hamiltonian systems defined by the vector fields $X_{h_i}$ is completely integrable.
\et  %

\br  %
This discussion has been ``formal'' in the sense that I have not considered global conditions, such as the completeness of the trajectories of $X_{h_j}$ within the surface $M_c$.  When $M_c$ is compact, then it is topologically a torus and it is possible to build action-angle variables.
See \cite{04-4} for a careful analysis of this.
\er  %

\subsubsection{Bi-Hamiltonian Systems}\label{sec:biham}

Within the context of soliton theory and the ``Lenard scheme'' the notion of {\em bi-Hamiltonian systems} was introduced in \cite{78-4,79-5} (see also \cite{86-1,fb90-4,93-8}). Here we just present some basic facts, which find use in the discussion of Section \ref{pn1-even}.

Let $P_1$ and $P_2$ define Poisson tensors (see Definition \ref{def:Ptensor}) on a given Poisson manifold.
\bd[Compatible Poisson Brackets]\label{compat-pb} %  %
The matrices $P_1$ and $P_2$ are said to be compatible Poisson tensors if $P_1$, $P_2$ and $P=P_2+P_1$ satisfiy the Jacobi identities (\ref{pdp}).  We can then define the following Poisson brackets:
$$
\{f,g\}_i = \nabla f P_i \nabla g , \quad i=1,2\quad\mbox{and}\quad
    \{f,g\}_P = \nabla f (P_2+P_1) \nabla g.
$$
\ed  %
Typically, both Poisson brackets will be degenerate, allowing us to define a {\em bi-Hamiltonian ladder}, starting with a Casimir function of $P_1$ and ending with one of $P_2$:
\be   \label{ladder}  %
P_1 \nabla h_1=0, \;\; P_1\nabla h_k = P_2 \nabla h_{k-1} ,
\;\;\;\mbox{for}\;\;\; 2\leq k\leq M,\;\;\;\mbox{and}\;\;\; P_2\nabla h_M=0,
\ee  %
for some $M$.  It is possible to start with the Casimir $h_1$ of $P_1$ and use the ladder relations to successively {\em construct} $h_k$ for $k>1$.  When we reach $h_M$ the ladder relations stop, with $P_2\nabla h_M=0$.

To prove complete integrability, we use the following result.
\bl[Bi-Hamiltonian Relations] \label{biham-rels} %
With the Poisson brackets given by Definition \ref{compat-pb}, the functions
$h_1, \dots , h_M$ satisfy
$$
\{h_i,h_j\}_1=\{h_i,h_{j-1}\}_2 \quad\mbox{and}\quad  \{h_i,h_j\}_2=\{h_{i+1},h_j\}_1 .
$$
\el  %
\begin{prf}
The ladder relations (\ref{ladder}) imply
$$
\{h_i,h_j\}_1 = \nabla h_i P_1 \nabla h_j = \nabla h_i P_2 \nabla h_{j-1} = \{h_i,h_{j-1}\}_2,
$$
and the second formula follows from this after writing $\{h_j,h_{i+1}\}_1=\{h_j,h_i\}_1$.
\end{prf}

\bt[Complete Integrability]\label{involut}   %
The functions $h_1, \dots , h_M$ are in involution with respect to both of the
above Poisson brackets
$$
\{h_i,h_j\}_1 = \{h_i,h_j\}_2 = 0,\quad\mbox{and hence}\quad \{h_i,h_j\}_P = 0.
$$
It then follows from Liouville's Theorem \ref{liouville} that the functions $h_1,\dots ,h_M$
define a completely integrable Hamiltonian system.
\et  %

\begin{prf}
Without loss of generality, choose $i<j$.  Then
$$
\{h_i,h_j\}_1=\{h_i,h_{j-1}\}_2= \{h_{i+1},h_{j-1}\}_1= \cdots =
\{h_k,h_k\}_\ell=0,
$$
for some $k,\ell$.  Similarly
$$
\{h_i,h_j\}_2=\{h_{i+1},h_j\}_1= \{h_{i+1},h_{j-1}\}_2= \cdots =
\{h_k,h_k\}_\ell=0,
$$
for some $k,\ell$.
\end{prf}

We can arrange the ladder relations to form
\be   \label{biham}  %
(P_2+P_1)(\nabla h_M -\nabla h_{M-1} +\nabla h_{M-2}-\dots
         +(-1)^{M+1} \nabla h_1)=0,
\ee   %
so
\be  \label{casimir}  %
{\cal C} = h_M - h_{M-1} + h_{M-2}-\dots +(-1)^{M+1} h_1
\ee  %
is the Casimir function of the Poisson matrix $P$.  Such a formula arises in our discussion of Section \ref{primit}.

\subsubsection{Super-Integrability of a Hamiltonian System}

On the symplectic leaves of dimension $2r$ we can have at most $r$ Poisson commuting functions (in which case we have complete integrability).  However, we can have up to $2r-1$ independent functions Poisson commuting with a single function $H$.  When we have $s$ independent functions $\{f_k, k=1,\dots s\}$, with $r<s\leq 2r-1$, then the system is said to be {\em super-integrable}.  When $s=2r-1$, $H$ is said to be {\em maximally super-integrable}.  In this case, the trajectories of $X_H$ must lie in the common level set of $2r-1$ functions.  Since the latter is itself $1-$dimensional, this common level set (determined by initial conditions) is just the trajectory itself and (in principle) is determined algebraically by solving the equations
$$
f_k=c_k,\quad  k=1,\dots , 2r-1.
$$
Well known examples of super-integrable systems are the {\em isotropic harmonic oscillator}, the {\em Kepler system} (additional integrals coming from the Runge-Lenz vector) and the Calogero-Moser system with inverse square potential.  The additional integrals are often called ``hidden symmetries''.

\subsection{Liouville Integrability of a Map}

We now consider a map $\varphi : M\rightarrow M$ in the context of Poisson manifolds.  We generalise the idea of a ``canonical transformation''.
We write
$$
(x_1,\dots ,x_n) \mapsto (\tilde x_1,\dots ,\tilde x_n),\quad\mbox{with}\quad \tilde x_i = \varphi_i(x_1,\dots ,x_n).
$$
Now suppose $P$ defines a Poisson bracket, with $\{x_i,x_j\}=P_{ij}$.  Then
$$
\{\tilde x_i,\tilde x_j\} = \{\varphi_i(x_1,\dots ,x_n),\varphi_j(x_1,\dots ,x_n)\}.
$$
The right hand side of this is a {\em function of $(x_1,\dots ,x_n)$}.

If it can be written explicitly in terms of the variables $(\tilde x_1,\dots ,\tilde x_n)$, then we can define
$$
\tilde P_{ij}=\{\tilde x_i,\tilde x_j\}.
$$
This means
$$
\tilde P = \left.\frac{\pa \tilde x}{\pa x} P \left(\frac{\pa \tilde x}{\pa x}\right)^T\right|_{\tilde x}.
$$

\bd[Invariant Poisson Bracket]  %
If $P_{ij}=\{x_i,x_j\}$ and $\tilde P_{ij}=\{\tilde x_i,\tilde x_j\}$ have the same functional form
of their respective arguments, then the Poisson bracket defined by $P$ is said to be
invariant under the action of the map.
\ed  %

\bd[Invariant Function]  %
A function $f:M\rightarrow R$ is said to be invariant under the map $\varphi:M\rightarrow M$, if $f\circ \varphi = f$.
\ed  %

The notion of complete integrability of a Poisson map was introduced in \cite{87-13,91-4}.

\bd[Complete Integrability of a Poisson Map]  %
Let $M$ be a Poisson manifold with coordinates $x_1,\dots ,x_{n}$ and Poisson matrix $P$, rank $2r$.
Suppose $P$ is invariant under the map $\varphi:M\rightarrow M$ and that the Casimir functions ${\cal C}_k, \, k=1,\dots m$ are themselves invariant:
$$
{\cal C}_k\circ \varphi = {\cal C}_k \quad\mbox{for each} \;\; k.
$$
Suppose further that there exist $r$ independent, invariant functions $h_i,\, i=1, \dots r$, which are {\em in involution} with respect to the Poisson bracket defined by $P$:
$$
\{h_i,h_j\}=0 \quad\mbox{for all}\;\; i,j.
$$
Then the map $\varphi$ is said to be Liouville integrable.
\ed  %

Since the Casimir functions are invariant, each orbit of the map is on one particular symplectic leaf $M_s$.
Since the functions $h_i,\, i=1, \dots r$, are invariant, the orbit is further restricted to lie on $M_c$, for
values of $c_i$ determined by initial conditions.
Since $\{h_i,h_j\}=0$, the vector fields $X_{h_i}$ commute.

Since the Poisson bracket, its Casimir functions and the functions $h_i$ are all invariant,
the map commutes with the continuous flows of $X_{h_i}$.  These continuous flows are therefore {\em continuous symmetries} of the map.

\br[Connected Components]  %
The manifolds $M_c$ are often {\em not connected}.
Whilst the above continuous flows must lie entirely within a single connected component,
the commuting map can jump between components.

Unlike the continuous case, we cannot think of the map as some combination of
``time-advance maps'' of the commuting flows.
\er  %

\br[Invariance of Casimir Functions]  %
If $P_{ij}=\{x_i,x_j\}$ is degenerate, its Casimir functions are not guaranteed to be invariant, so the trajectory of the map may not lie within a symplectic leaf, in which case we cannot deduce Liouville integrability.  See Example \ref{somos4pbex} below.
\er  %

\br[Super-integrability]  %
Each symplectic leaf $M_s$ has dimension $2r$.  We have seen that there exist super-integrable {\em continuous} systems with at most $2r-1$ first integrals, since the common level set is already $1-$dimensional and hence just the (unparameterised) trajectory of $X_H$.  We can also have super-integrable {\em discrete} systems, but in this case it is even possible to have $2r$ first integrals, whose common set consists of a {\em finite} number of points.  In such cases the map is periodic.  Conversely, given a periodic map $\varphi$ (period $p$), any function $f({\bf x})$ will generate a finite sequence of functions $f({\bf x}), f(\varphi({\bf x})), \dots , f(\varphi^{p-1}({\bf x}))$, any cyclically symmetric combination of which represents a first integral.  In this way we can construct $2r$ independent first integrals.
\er  %

\subsection{Log-Canonical Brackets for Cluster Mutations}\label{gsv}

In \cite{03-5} it was shown that very general cluster algebras admit a linear space of Poisson brackets of log-canonical
type, compatible with the cluster maps generated by mutations, and having the form
\be\label{logcan}
\{x_j,x_k\}=c_{jk}\, x_{j}x_k,
\ee
for some skew-symmetric constant coefficient matrix $C=(c_{jk})$.
Compatibility of the Poisson structure means that the cluster transformations $\mu_i$ given by (\ref{ex-rel}) correspond
to a {\em change of coordinates}, $\tilde{\bf{ x}}=\mu_i ({\bf x}) $, with their bracket also being log-canonical,
$$
\{ \tilde{x}_j,\tilde{x}_k\}=\tilde{c}_{jk}\,\tilde{x}_{j}\tilde{x}_k,
$$
for another skew-symmetric constant matrix $\tilde{C}=(\tilde{c}_{jk})$.

If we regard the cluster transformation as a {\em birational map}
${\bf x}\mapsto \tilde{\bf{ x}}=\varphi ({\bf x})$ in $\C^N$, and require a Poisson
structure that is {\it invariant} with respect to $\varphi$ (not just {\em covariant}),
then we require $\tilde{C}=C$.  However, there may not be a non-trivial log-canonical Poisson bracket that is covariant or invariant under
cluster transformations.

\bex \label{affineA2}  {\em
Corresponding to the $3$ node period $1$ primitive $P_3^{(1)}$, the matrix $B$ and birational map on $\C^3$ are given by
\be \label{affA2}
B=\left(\begin{array}{ccc}
0 & -1 & -1 \\
1 & 0 & -1 \\
1 & 1 & 0 \end{array}\right),\qquad
\left(\begin{array}{c}
x_1 \\
x_2 \\
x_{3}
\end{array} \right)
\longmapsto \left(\begin{array}{c}
x_2 \\
x_3 \\
(x_{2}x_3+1)/x_1
\end{array} \right).
\ee
It is easy to see that there is no invariant Poisson bracket of the form  (\ref{logcan}).  However, we will see that it is possible to reduce (\ref{affA2}) to a map acting on a $2-$dimensional space on which there {\em does} exist a log-canonical bracket.
}\eex

Even when it is possible to find such a Poisson bracket it may not be the ``correct bracket'' for discussing complete integrability, since we require that any Casimir functions be {\em invariant} under the map.

\bex[Invariant Poisson Bracket for the Somos-$4$ Map] \label{somos4pbex} {\em  %
The Somos-$4$ map is
\be \label{birs4}
\varphi: \quad
\left(\begin{array}{c}
x_1 \\
x_2 \\
x_3 \\
x_4
\end{array} \right)
\longmapsto \left(\begin{array}{c}
x_2 \\
x_3 \\
x_4  \\
\displaystyle \frac{x_2x_4+x_3^2}{x_1}
\end{array} \right),
\ee
which preserves the log-canonical Poisson bracket (see \cite{07-2})
\be \label{logcans4}  %
\{x_j,x_k\}=c_{jk}\, x_{j}x_k,\quad\mbox{where}\quad  C=\left(\begin{array}{ccccc}
                                                         0 & 1 & 2 & 3 \\
                                                         -1 & 0 & 1 & 2 \\
                                                         -2 & -1 & 0 & 1 \\
                                                         -3 & -2 & -1 & 0
                                                         \end{array} \right).
\ee %
This has rank 2, with two independent null vectors
$$
{\bf m}_1=(1,-2,1,0)^T, \qquad
{\bf m}_2=(0,1,-2,1)^T,
$$
providing two independent Casimir functions for the bracket:
$$
y_1 = \frac{x_1 x_3}{x_2^2}, \qquad
y_2 = \frac{x_2 x_4}{x_3^2}.
$$
The map (\ref{birs4}) induces the following map on the $2-$dimensional $y-$space:
\be \label{s4map}
\hat{\varphi}: \quad
\left(\begin{array}{c}
y_1 \\
y_2
\end{array} \right)
\longmapsto \left(\begin{array}{c}
y_2 \\
(y_2+1)/(y_1y_2^2)
\end{array} \right),
\ee
which is of QRT type \cite{88-5} (a well known class of integrable maps), and preserves the log-canonical bracket
\be\label{canon}
\{y_1,y_2\}=y_1y_2.
\ee
A similar result holds for the Somos-$5$ map (see \cite{f11-3}).
}\eex %

The finding of Poisson brackets such as (\ref{logcans4}) is ad-hoc, by assuming a bracket of general form (\ref{logcan}) and using the map to fix the values of $c_{ij}$ ({\em if possible!}).  This bracket can turn out to be unsuitable, with non-invariant symplectic leaves.  In the case of Somos-$4$, the associated Liouville integrable map is not (\ref{birs4}) at all, but the induced map on the Casimir functions, for which another Poisson bracket had to be found by ad-hoc calculations.

In fact, it is possible to systematically construct all these Poisson brackets from a knowledge of the matrix $B$, which defines the original periodic quiver.  In the next section we consider an invariant two-form in the variables $x_j$, which naturally exists for any cluster map.

\section{Symplectic Maps from Cluster Recurrences}\label{symplectic}
\setcounter{equation}{0}

Given a recurrence, a major problem is to find an appropriate
symplectic or Poisson structure which is invariant under the action of the corresponding
finite-dimensional map.  Remarkably, in the case of cluster recurrences (from mutation periodic quivers)
this problem can be solved algorithmically.

For the period $m$ case, we need to combine the sequence of $m$ mutations to form a map which preserves $B(1)$.
Most of the discussion below is concerned with the period $1$ case, but some period $2$ examples are presented.

\subsection{Maps from Period $1$ Recurrences}

Given a period $1$ quiver, with matrix $B$, the cluster recurrence (\ref{arec}) corresponds to the map $\varphi:\C^N\rightarrow \C^N$, given by
\be \label{bir}
\varphi: \quad
\left(\begin{array}{c}
x_1 \\
x_2 \\
\vdots  \\
x_{N-1} \\
x_{N}
\end{array} \right)
\longmapsto \left(\begin{array}{c}
x_2 \\
x_3 \\
\vdots  \\
x_{N} \\
x_{N+1}
\end{array} \right),
\qquad \mathrm{where} \qquad
 x_{N+1}=\frac{
\prod_{m_j\geq 0}x_{j+1}^{m_j}+ \prod_{m_j\leq 0}x_{j+1}^{-m_j}}{x_1}.
\ee
One can decompose the map (\ref{bir}) as $\varphi =\rho^{-1}\circ \mu_1$, where in
terms of the cluster ${\bf x}=(x_j)$ the map $\mu_1$ sends
$(x_1, x_2, \ldots , x_N)$ to $(\tilde{x}_1, x_2,    \ldots , x_N)$, with
$\tilde{x}_1$ defined according to the exchange relation (\ref{ex-rel})
with $k=1$, and $\rho^{-1}$ sends $(x_1, x_2, \dots , x_N)$ to $( x_2,\dots , x_N, x_1)$.
We have already seen the examples (\ref{affA2}) and (\ref{birs4}).

Given the same skew-symmetric matrix $B$, one can define the log-canonical two-form
\be \label{omega}
\omega =\sum_{j<k} \frac{b_{jk}}{x_j x_k} d x_j\wedge d x_k,
\ee
which is just the constant skew-form $\omega = \sum_{j<k} b_{jk}\, d z_j\wedge d z_k$, written in the logarithmic coordinates $z_j=\log x_j$,
so it is evidently closed, but may be degenerate.  Such a form is called {\em pre-symplectic} and is a genuine {\em symplectic form} when $B$ is non-degenerate.

\br[Covariance vs Invariance]    %
In  \cite{05-6} it was shown that for a cluster algebra defined by a skew-symmetric integer matrix $B$,
this two-form is compatible with cluster transformations, in the sense that
under a mutation map $\mu_i : \, {\bf x} \mapsto \tilde{ {\bf x}}$, it transforms as
$\mu_i^* \omega =\sum_{j<k}\tilde{b}_{jk} d\log\tilde{x}_j\wedge d\log\tilde{x}_k$.  It is important for us that for a period $1$ quiver, $\tilde B=B$, so this $2-$form is \underline{invariant}.
\er  %

\bl \label{invar-symp}
Let $B$ be a skew-symmetric integer matrix, corresponding to a cluster mutation-periodic quiver with period 1.

Then the two-form $\omega$ is preserved by the map $\varphi$, i.e. $\varphi^* \omega =\omega$.
\el
When $B$ is non-singular, we define $C$ to be any convenient integer matrix proportional to $B^{-1}$ and use this to define a log-canonical Poisson bracket (\ref{logcan}).  Otherwise we must first reduce $\omega$ to a nonsingular form on its symplectic leaves.

\bex[The $4$ Node Period $1$ Quiver]\label{s4gen-ex} {\em
For integer values of $c\geq 0$, the skew-symmetric matrix
\be \label{s4gen}
B=\left(\begin{array}{cccc} 0 & -1 & c & -1 \\
                    1 & 0  & -(c+1) & c \\
                    -c & (c+1) & 0 & -1 \\
                     1 & -c & 1 & 0
\end{array}\right),
\ee
defines a period 1 cluster mutation-periodic quiver $Q$, giving rise to the recurrence $x_{n+4}\, x_{n} = x_{n+3}\, x_{n+1}+x_{n+2}^c$.
Thus, by Lemma \ref{invar-symp}, for each $c$ the map $\varphi$, given by
\be \label{bir4node-c}
\varphi: \quad
\left(\begin{array}{c}
x_1 \\
x_2 \\
x_3 \\
x_4
\end{array} \right)
\longmapsto \left(\begin{array}{c}
x_2 \\
x_3 \\
x_4  \\
\displaystyle \frac{x_2x_4+x_3^c}{x_1}
\end{array} \right),
\ee
preserves the two-form
\be \label{symp4node-c}
\omega =
  -\left( \frac{d x_1 \wedge d x_2}{x_1x_2}+\frac{d x_1 \wedge d x_4}{x_1x_4}+
\frac{d x_3 \wedge d x_4}{x_3x_4}\right)
+c \left(\frac{d x_1 \wedge d x_3}{x_1x_3}+ \frac{d x_2 \wedge d x_4}{x_2x_4}\right)
-(c+1) \frac{d x_2 \wedge d x_3}{x_2x_3}.
\ee
For $c\neq 2$ this is symplectic ($B$ non-singular) and leads to the log-canonical bracket with
$$
C=\left(
    \begin{array}{cccc}
      0 & 1 & c & 1+c \\
      -1 & 0 & 1 & c \\
      -c & -1 & 0 & 1 \\
      -1-c & -c & -1 & 0
    \end{array}
  \right).
$$
Even when $c=2$ this matrix defines an invariant Poisson bracket for (\ref{bir4node-c}) and is identical to (\ref{logcans4}).
}\eex

\subsubsection{Reduction to Symplectic Leaves}

If $B$ is singular (of rank $2r$, with $N=2r+m$) we consider the null distribution of $\omega$, which (away from
the hyperplanes $x_j=0$) is generated by $m$ independent commuting vector fields, each of which is of the form
\be\label{null}
{\bf X_u} = \sum_{j=1}^N u_j x_j \, \frac{\pa}{\pa x_j}
\qquad \mathrm{for} \qquad {\bf u}=(u_j)\in\mathrm{ker} \, B.
\ee
Since this is an integrable distribution, Frobenius' theorem gives local coordinates
$t_1, \ldots , t_{m} , y_1,\ldots ,y_{2r}$ such that the integral manifolds of the null distribution
are given by $y_j = \mbox{constant},\, j=1,\ldots ,2r$.
The coordinates $y_j$ must be common invariants for these commuting vector fields, and can be chosen as linear functions of
the logarithmic coordinates $z_j = \log x_j$, so each $y$ is of the form
$$
y= {\bf x}^{\bf v} :=\prod_j x_j^{v_j} \qquad \iff \qquad ({\bf u}, {\bf v})=0, \quad \forall \, {\bf u}\in \mathrm{ker}\, B ,
$$
where $(\, , )$ denotes the standard scalar product.

On the symplectic leaves, with coordinates $y_j$, $\omega$ defines a log-canonical symplectic form
\be\label{yform}
\hat{\omega} = \sum_{j<k} \frac{\hat{b}_{jk}}{y_j y_k}d y_j\wedge d y_k.
\ee

\bex [Somos-$4$]\label{s4e} {\em
Returning to example (\ref{s4gen-ex}), with $c=2$, $\mathrm{ker}\, B$ is spanned by the integer column vectors
$\mathbf{u}_1= (1,1,1,1)^T , \;\; \mathbf{u}_2= (1,2,3,4)^T$.
Then $\mathrm{im}\, B=(\mathrm{ker}\, B)^\perp=\; <{\bf v}_1, {\bf v}_2>$, with
${\bf v}_1=(1,-2,1,0)^T, \;\; {\bf v}_2=(0,1,-2,1)^T$,
whose components provide the exponents for the monomial invariants
\be\label{s4yj}
y_1 = \frac{x_1x_3}{x_2^2}, \qquad y_2 = \frac{x_2x_4}{x_3^2}.
\ee
(These monomials are just two independent Casimirs for the degenerate
Poisson bracket (\ref{logcans4}), which can, in fact, be written as the bivector ${\bf u}_1\wedge {\bf u}_2$.)

As we already saw, the map (\ref{birs4}) reduces to the map (\ref{s4map}), which is of QRT type \cite{88-5}, and preserves the symplectic form
\be\label{s4canon}
\hat\omega = \frac{1}{y_1y_2}\, d y_2\wedge d y_1,
\ee
whose ``inverse'' defines the Poisson bracket (\ref{canon}).
}\eex

For a given example it is straightforward to calculate coordinates $y_i$, such that the $y-$map is birational.
In \cite{f14-1} we proved the following theorem, which guarantees the existence of such a birational map.
\bt \label{torusred}
The map $\varphi$ is symplectic whenever $B$ is nondegenerate.
%In the degenerate case that
For $\mathrm{rank} \, B = 2r \leq N$,
there is a rational map $\pi$ and a symplectic
birational map $\hat\varphi$ such that the diagram
\be \label{cd}
\begin{CD}
\C^N @>\varphi >> \C^N\\
@VV\pi V @VV\pi V\\
\C^{2r} @>\hat{\varphi}>> \C^{2r}
\end{CD}
\ee
is commutative, with a log-canonical symplectic form $\hat\omega$ on $\C^{2r}$ that satisfies
$\pi^*\hat\omega = \omega$.
\et

\subsection{Maps from Period $2$ Recurrences}

For the period $2$ case, we need to combine the sequence of $2$ mutations to form a map which preserves $B(1)$.

\bex[The $4$ Node Case]   {\em
We take the quivers of Example \ref{reg4}.
Recall that we alternate the labels of the cluster variables, with $x_n,\, y_n,\, x_{n+1},\, y_{n+1}$ at nodes $1-4$, respectively.
After $2$ mutations we have the recurrence (\ref{period2-4xy}), which leads to the map
$$
(x_1,y_1,x_2,y_2)\mapsto (x_2,y_2,\tilde x_2,\tilde y_2),\quad\mbox{where}\quad
     \tilde x_2=\frac{y_1^ry_2^t+x_2^s}{x_1},\;\;  \tilde y_2 = \frac{x_2^t \tilde x_2^r+y_2^s}{y_2}.
$$
Hence this map has the invariant (pre-)symplectic form (\ref{omega}) with matrix $B(1)$, taking the explicit form
$$
\omega = r \frac{d x_1\wedge d y_1}{x_1y_1}- s \frac{d x_1\wedge d x_2}{x_1x_2} +t \frac{d x_1\wedge d y_2}{x_1y_2}
      +t \frac{d x_2\wedge d y_1}{x_2y_1} - (r+s t) \frac{d x_2\wedge d y_2}{x_2y_2}+s \frac{d y_1\wedge d y_2}{y_1y_2},
$$
which is degenerate if and only if $r^2-s^2+t^2+r s t=0$.
}\eex   %

\bex[The $5$ Node Case]   {\em
We take the quivers of Example \ref{reg5} (with $r=1, t=0$ for simplicity),
with initial cluster variables $x_1,\, y_1,\, x_2,\, y_2,\, x_3$ at nodes $1-5$, respectively.
After $2$ mutations we obtain the $5-$dimensional map
$$
(x_1,y_1,x_2,y_2,x_3)\mapsto \left(x_2,y_2,x_3,\tilde y_2,\tilde x_3\right),\quad\mbox{where}\quad
     \tilde y_2=\frac{y_1+x_2y_2}{x_1},\;\;  \tilde x_3 = \frac{y_1+x_2y_2+x_1x_3y_2}{x_1y_1}.
$$
This map has the invariant (pre-)symplectic form (\ref{omega}) with matrix $B(1)$, taking the explicit form
$$
\omega =  \frac{d x_1\wedge d x_2}{x_1x_2}-\frac{d x_1\wedge d y_1}{x_1y_1} +\frac{d x_1\wedge d y_2}{x_1y_2}
      +\frac{d x_2\wedge d y_1}{x_2y_1} -  \frac{d x_2\wedge d y_2}{x_2y_2}
      +\frac{d x_2\wedge d x_3}{x_2x_3}- \frac{d x_3\wedge d y_1}{x_3y_1}+ \frac{d x_3\wedge d y_2}{x_3y_2},
$$
which is degenerate with null vector $(0,y_1,0,y_2,0)$.  On symplectic leaves we may choose
$$
q_1=x_1,q_2=x_2,q_3=x_3,q_4= \frac{y_1}{y_2}, \quad\mbox{giving}\quad (q_1,q_2,q_3,q_4)\mapsto\left(q_2,q_3,\frac{q_2+q_1q_3+q_4}{q_1q_4},\frac{q_1}{q_2+q_4}\right)
$$
and
$$
\omega = \frac{d q_1\wedge d q_2}{q_1q_2}-\frac{d q_1\wedge d q_4}{q_1q_4}+\frac{d q_2\wedge d q_3}{q_2q_3}+
           \frac{d q_2\wedge d q_4}{q_2q_4}-\frac{d q_3\wedge d q_4}{q_3q_4}.
$$
}\eex   %

\section{Integrability of Cluster Maps from Periodic Quivers}\label{integrability}
\setcounter{equation}{0}

We have seen that each periodic quiver gives rise to a recurrence with the Laurent property.  Furthermore, we can write our recurrences as maps on finite dimensional spaces (a space of dimension $N$ for a period $1$ quiver with $N$ nodes) and these maps possess an invariant pre-symplectic form, given purely in terms of the matrix $B$ which defines the quiver.  All of this is algorithmic.

The next question is much more difficult to answer.  {\em Which of these maps is completely integrable?}  We need to first isolate those cases which have some chance of being integrable.  We can use one of the ``tests for integrability'', such as calculating the {\em algebraic entropy}.
I give a brief description of this for period $1$ recurrences.

We then need to analyse those maps which ``pass'' this test.  Our most important task is to construct the ``correct number'' of Poisson commuting invariant functions on the symplectic leaves of $\omega$.

\subsection{Algebraic Entropy and Tropical Recurrences} \label{entropy}

The connection between the integrability of maps and various weak growth properties of the iterates has been appreciated for some time
(see \cite{91-4} and references).  In the case of rational maps, Bellon and Viallet \cite{99-12}
considered the growth of degrees of iterates, and used this to define a notion of
algebraic entropy. Each component of a rational map $\varphi$ in affine space
is a rational function of the coordinates, and the degree of the map, $d=\mathrm{deg}\,\varphi$,  is the maximum
of the degrees of the components.  By iterating the map $n$ times one gets a sequence of rational functions
whose degrees grow generically like $d^n$. At the $n$th step one can set $d_n=\mathrm{deg}\,\varphi^n$,
and then the algebraic entropy $\mathcal{E}$ of the map is defined to be
$
\mathcal{E}=\mathrm{lim}_{n\to\infty}\frac{1}{n}\log d_n.
$
Generically, for a map of degree $d$, the entropy is $\log d>0$, but for special maps there can be
cancellations in the rational functions that appear under iteration, which means that the entropy is smaller than expected.

It is conjectured that Liouville-Arnold integrability corresponds to zero algebraic entropy.
In an algebro-geometric setting, there are plausible
arguments which indicate that zero entropy should be a necessary condition for
integrability in the Liouville-Arnold sense. In the latter setting, each iteration of the map
corresponds to a translation on an Abelian variety (the level set of the first integrals),
and the degree is a logarithmic height function, which necessarily grows like
$d_n\sim \mathrm{C}n^2$.

Often the algebraic entropy of a map can only be guessed experimentally, by calculating the degree sequence $(d_n)$
up to some large $n$ and doing a numerical fit to a linear recurrence. This is increasingly impractical as the dimension increases, and provides no proof that the linear relation, with its corresponding entropy value, is correct.

The Laurent property for period $1$ recurrences implies that the iterates have the factorised form
$$
x_n = \frac{\mathrm{N}_n({\bf x})} {\mathrm{M}_n({\bf x})},
\qquad \mathrm{with} \quad \mathrm{N}_n \in\Z [ {\bf x} ] = \Z [x_1,\ldots ,x_N], \quad
\mathrm{M}_n =\prod_{j=1}^N x_j^{d^{(j)}_n},
$$
where the  polynomials $\mathrm{N}_n$ are not divisible by $x_j$ for $1\leq j\leq N$, and $\mathrm{M}_n$ are Laurent monomials, which leads to
the dynamics of the $\mathrm{M}_n$ being decoupled from the $\mathrm{N}_n$.

This leads to a tropical (or ultradiscrete \cite{08-6}) analogue of the original nonlinear recurrence (\ref{arec}), in terms of the max-plus algebra.
\bp \label{troprec}
For all $n$, the exponent $d^{(j)}_n$ of each variable in the Laurent monomial $\mathrm{M}_n$ satisfies the recurrence
\be \label{tropical}
d_{n+N}+d_n = \max \left( \, \sum_{j=1}^{N-1} [b_{1,j+1}]_+\, d_{n+j}\, , \, \sum_{j=1}^{N-1}  [-b_{1,j+1}]_+ \,d_{n+j}\, \right) ,
\ee
with the initial conditions $d_1=-1$, $d_2=\ldots =d_N=0$ (up to shifting the index).
\ep

\bex[Tropical Somos-$4$] \label{ts4}  {\em
The tropical version of the Somos-$4$ recurrence  is
\be\label{tsomos4}
d_{n+4}+d_{n}=\max (d_{n+3}+d_{n+1},2d_{n+2}).
\ee
With initial conditions $d_1=-1$ and $d_2=d_3=d_4=0$ this  generates a sequence that begins
$$
-1,0,0,0,1,1,2,3,3,5,6,7,9,10,12,14,15,18,20,22,25,27,30,33,\ldots ,
$$
which are the degrees (in each of the variables $x_1,x_2,x_3,x_4$) of the denominators of the Laurent polynomials
generated by (\ref{somos4}). The preceding sequence has quadratic growth, $d_n\sim \mathrm{C}n^2$ as $n\to\infty$ (consistent with the growth
of logarithmic height on an elliptic curve), so that the algebraic entropy is zero.
}\eex  %

Using (\ref{tropical}), together with a number of conjectures from the cluster algebra literature (see \cite{f14-1} for details), we arrive at the following conjecture.
\bcon \label{entropyconj}
For a birational map given by  (\ref{bir}), corresponding to a recurrence of the form (\ref{arec}), the algebraic entropy is
positive if and only if
\be \label{maxe}
\max \left(\,  \sum_{m_j>0} m_j \, , \, \sum_{m_j<0} (-m_j) \, \right) \geq 3,
\ee
\econ

\bt\label{zeroe}
Suppose that Conjecture \ref{entropyconj} holds. Then there are only four
distinct cases for which the algebraic entropy is zero, giving the following families of
recurrences:

\smallskip\noindent {\em (i)} For even $N=2m$ only, we have the recurrence
\be\label{per5}
x_{n+2m}\, x_n = x_{n+m} +1.
\ee

\noindent {\em (ii)} For each $N\geq 2$ and $1\leq q\leq\lfloor{N/2}\rfloor$, we have
the recurrence
\be \label{primrec}
x_{n+N}\, x_n = x_{n+N-q}\, x_{n+q} + 1.
\ee

\noindent {\em (iii)}  For even $N=2m$ only, and  $1\leq q\leq m-1$, we have the recurrence
\be \label{comprec}
x_{n+2m}\, x_n = x_{n+2m-q}\, x_{n+q} + x_{n+m}.
\ee

\noindent {\em (iv)} For each $N\geq 2$ and $1\leq p <q\leq \lfloor{N/2}\rfloor$, we have the recurrence
\be \label{somosN}
x_{n+N}\, x_n = x_{n+N-p}\, x_{n+p} + x_{n+N-q}\, x_{n+q}.
\ee
\et
The simplest is case (i), where the recurrence (\ref{per5}) decouples into $m$ independent copies of the Lyness map: all the orbits
are periodic, and the overall period of the sequence of $x_n$ is $5m$.

The families (ii) and (iii) are linearisable and super-integrable and discussed in Sections \ref{primit} and \ref{pert}.
The family (iv) includes the Somos-$4$ and $5$ recurrences as well as the $3-$term Gale-Robinson recurrences and will be discussed in Section \ref{grmaps}.

This Theorem only concerns period $1$ recurrences.  There is currently no classification of integrable cases of recurrences associated with higher periodicity.

\section{Linearisable Recurrences from Period $1$ Primitives}\label{primit}
\setcounter{equation}{0}

We already saw in Theorem \ref{p1-theorem} that the general period 1 quiver is built out of period $1$ primitives.  It turns out that the period $1$ primitives themselves give rise to interesting recurrences.  They correspond to case (ii) of Theorem \ref{zeroe} and can be rewritten as
\be
x_{n+N}\, x_n = x_{n+p}\, x_{n+q} + 1, \qquad p+q=N.
\label{jkrec}
\ee
When $p$ and $q$ are coprime, the associated quivers can be ``unfolded'' to be mutation-equivalent to
the affine Dynkin quivers $\tilde{A}_{q,p}$,
corresponding to the $A_{N-1}^{(1)}$ Dynkin diagram with $q$ edges oriented clockwise and $p$ oriented
anticlockwise, while if
$\gcd (p,q)=k>1$ so that $p=k\hat{p}$, $q=k\hat{q}$, then the quiver is just the disjoint union of $k$ copies
of  $\tilde{A}_{\hat{q},\hat{p}}$.  Whilst these quivers are of {\em finite mutation type} the corresponding cluster algebras are not finite and the recurrences generate infinite sequences.

Much of the general structure will be discussed for arbitrary $q$, but a detailed description of complete integrability of the associated maps (Sections \ref{pn1-even} and \ref{pn1-odd}) is restricted to the case $q=1$.  The key to understanding the complete integrability is the structure of the Poisson algebras generated by some functions $J_n$ and $K_n$, defined by Equations (\ref{Jrec}) and (\ref{Krec}) below.  The algebras for the case $q\neq 1$ can be shown to be isomorphic to others for $q=1$, but the technical details are fairly complicated (see \cite{f14-1}).

\subsection{Linear Relations with Periodic Coefficients}

We can write the nonlinear recurrence (\ref{jkrec}) in the form
\be \label{frieze}
\det \,\Psi_n = 1,\quad\mbox{where}\quad
           \Psi_n = \left( \begin{array}{cc} x_n & x_{n+q} \\
                                 x_{n+p} & x_{n+N}
\end{array}\right).
\ee
Upon forming the matrix
\be\label{3by3}
 \tilde{\Psi}_n = \left( \begin{array}{ccc} x_n & x_{n+q} & x_{n+2q} \\
                                 x_{n+p} & x_{n+N}  & x_{n+N+q} \\
                                 x_{n+2p} & x_{n+N+p}  & x_{n+2N}
\end{array}\right),
\ee
one can use Dodgson condensation \cite{1866-1} to expand the $3\times 3$ determinant
in terms of its $2\times 2$ minors, as
$$
\det\, \tilde{\Psi}_{n}=\frac{1}{x_{n+N}}\left( \det\Psi_n \, \det \Psi_{n+N}-\det\Psi_{n+q} \, \det \Psi_{n+p}\right)=0,
$$
by (\ref{frieze}). By considering the right and left kernels of $\tilde{\Psi}_{n}$, we are led to the following result.
\bl \label{JK}
The iterates of the recurrence (\ref{jkrec}) satisfy the linear relations
\bea
&& x_{n+2q}-J_n\, x_{n+q} + x_n = 0,  \label{Jrec}  \\
&& x_{n+2p}-K_n\, x_{n+p} + x_n = 0,  \label{Krec}
\eea
whose coefficients are periodic functions of period $p,q$ respectively, that is
$$
J_{n+p}=J_n, \qquad K_{n+q}=K_n, \qquad  for\,\,  all \, \, n.
$$
In particular, when $q=1$ we have $K_{n+1}=K_n=\mathcal{K}$, constant for all $n$, so $x_n$ satisfies a {\em constant coefficient, linear difference equation}
\be\label{k-pn1}  %
x_{n+2N-2}+x_n=\mathcal{K} \, x_{n+N-1}.
\ee
\el
\br  %
The key idea is that since $\tilde{\Psi}_n$ and $\tilde{\Psi}_{n+p}$ share two rows, they have the same right null vector.  Similarly $\tilde{\Psi}_n$ and $\tilde{\Psi}_{n+q}$ share two columns.
\er  %

We can use the recurrence (\ref{jkrec}) to write $J_1, \dots , J_p, K_1,\dots K_q$ in terms of initial values $x_1, \dots , x_N$ and then any cyclically symmetric combination of the $J_i$ or of the $K_i$ furnishes us with a first integral for the map $\varphi$ derived from (\ref{jkrec}).  Alternatively, starting with
$$
J_1=\frac{x_1+x_{2q+1}}{x_{q+1}},
$$
written in terms of $x_1, \dots , x_N$, we can use the map $\varphi$ to generate independent functions $J_2, \dots , J_p$.  Similarly, we can generate $K_2, \dots , K_q$ from $K_1$.  Of course, we cannot have $p+q=N$ independent integrals, since these maps are not periodic.  From the monodromy relations derived from (\ref{Jrec}) and (\ref{Krec}), we can derive a single polynomial relationship between the functions $J_i$ and $K_i$ (see (\ref{k2})).  The existence of $N-1$ independent first integrals for $\varphi$ means that it is maximally super-integrable.

\subsection{Monodromy Matrices and Linear Relations with Constant Coefficients}\label{2x2monodromy}

The relation (\ref{Jrec}) implies that the matrix $\Psi_{n}$ satisfies
\be\label{psirel1}
\Psi_{n+q}=\Psi_n\, {\bf L}_n, \qquad {\bf L}_n=\left(\begin{array}{cc}
                                                   0 &  -1 \\
                                                   1  & J_n
                                                   \end{array}\right).
\ee
Upon taking the ordered product of the ${\bf L}_n$ over $p$ steps, shifting by $q$ each time, we have
the monodromy matrix
\be\label{mdy1}
{\bf M}_n:=  {\bf L}_{n}{\bf L}_{n+q}\ldots{\bf L}_{n+(p-1)q}=\Psi_n^{-1}\, \Psi_{n+pq}.
\ee
On the other hand, the recurrence (\ref{Krec}) yields
\be\label{psirel2}
\Psi_{n+p}= \hat{\bf L}_n \,\Psi_n, \qquad \hat{\bf L}_n=\left(\begin{array}{cc}
                                                                 0 &  1 \\
                                                                 -1  & K_n
                                                                 \end{array}\right),
\ee
which gives another monodromy matrix
\be\label{mdy2}
\hat{\bf M}_n:=  \hat{\bf L}_{n+(q-1)p}\ldots\hat{\bf L}_{n+p}\hat{\bf L}_n=\Psi_{n+pq}\, \Psi_n^{-1}.
\ee
The cyclic property of the trace implies that
\be\label{k2}
\mathcal{K}_n:= \mathrm{tr}\, {\bf M}_n = \mathrm{tr}\, \hat{\bf M}_n.
\ee
Also, since ${\bf L}_n$ has period $p$, shifting $n\to n+p$ in (\ref{mdy1}) and taking the trace implies that
$\mathcal{K}_{n+p}=\mathcal{K}_n$. Similarly, from (\ref{mdy2}) we have $\mathcal{K}_{n+q}=\mathcal{K}_n$.
Since the periods $p$ and $q$ are coprime it follows that $\mathcal{K}_n=\mathcal{K}= \mbox{constant}$, for all $n$, so $\mathcal{K}$ is an invariant of $\varphi$.

We can now generalise the result (\ref{k-pn1}).  This is Theorem 9.5 of \cite{f11-2}, but this particular proof is from \cite{f14-1}.
\bt\label{Klin}
The iterates of the nonlinear recurrence (\ref{jkrec}) satisfy the linear relation
\be\label{krec}
x_{n+2pq}+x_n = \mathcal{K}\,x_{n+pq},
\ee
where  $\mathcal{K}$ is the first integral defined by (\ref{k2}),
with $\mathcal{K}_n=\mathcal{K}$ for all $n$.
\et
\begin{prf}
Using (\ref{mdy1}) we see that $\Psi_{n+pq}=\Psi_n {\bf M}_n$, so $\Psi_{n+2pq}=\Psi_{n}{\bf M}_n{\bf M}_{n+pq}=\Psi_n {\bf M}_n^2$
by periodicity.  Noting that ${\bf M}_n$ is a $2\times 2$ matrix, with
$\det {\bf M}_n=1$ and $\mathrm{tr}\, {\bf M}_n={\cal K}$, yields
(by Cayley-Hamilton)
$$
\Psi_{n+2pq}-{\cal K} \Psi_{n+pq}+\Psi_n = \Psi_n({\bf M}_n^2-{\cal K} {\bf M}_n+{\bf I})=0.
$$
The $(1,1)$ component of this equation is just (\ref{krec}).
\end{prf}

\subsection{The Structure of Monodromy Matrices}

Since the only non-trivial invariant of ${\bf M}_n$ (or indeed $\hat {\bf M}_n$) is ${\cal K}$, it is possible to write the whole matrix ${\bf M}_n$ in terms of ${\cal K}$ and its derivatives.  We just present the case of $P_{N}^{(1)}$.

For the case $q=1$, we denote the matrix ${\bf M}_n={\bf L}_n {\bf L}_{n+1}\cdots {\bf L}_{n+p-1}$
by ${\bf M}_n^{(2m)}$, when $p=2m-1, m\geq 1$, or ${\bf M}_n^{(2m+1)}$, when $p=2m, m\geq 1$.

It is important to note that ${\bf M}_n^{(p+1)}$ depends only upon the variables $J_n,\dots ,J_{n+p-1}$.
For the moment, this is not really a monodromy matrix, if we do not assume any periodicity.
We give a recursive procedure for building the matrices ${\bf M}_n^{(2m)}$ and ${\bf M}_n^{(2m+1)}$.

We use the recursion ${\bf M}_n^{(p+3)}={\bf M}_n^{(p+1)}{\bf L}_{n+p}{\bf L}_{n+p+1}$ to prove the following.
\bl[The structure of ${\bf M}_n$]  \label{brelns}%
The components of ${\bf M}_n^{(p+1)}= \left(
                                    \begin{array}{cc} \RA^{(p+1)}_n & \RB^{(p+1)}_n \\
                                                \RC^{(p+1)}_n & \RD^{(p+1)}_n
                                          \end{array}\right)$ satisfy the relations
$$
\frac{\pa \RA_n^{(p+1)}}{\pa J_n}=\frac{\pa \RB_n^{(p+1)}}{\pa J_n}=0,\quad \RA_n^{(p+1)}= -\frac{\pa \RC_n^{(p+1)}}{\pa J_n}, \quad
\RB_n^{(p+1)}=-\frac{\pa {\cal K}^{(p+1)}}{\pa J_n},\quad \RC_n^{(p+1)}=\frac{\pa {\cal K}^{(p+1)}}{\pa J_{n+p-1}},
$$
where ${\cal K}^{(p+1)}=\RA_n^{(p+1)}+\RD_n^{(p+1)}$.

Furthermore, we can build the functions ${\cal K}^{(p+1)}$ by use of a {\em recursion operator}.  We have ${\cal K}^{(p+3)}={\cal R}^{(p)}{\cal K}^{(p+1)}$, where the {\em recursion operator} is
\be\label{krecursion} %
{\cal R}^{(p)} = J_{n+p}J_{n+p+1}\frac{\pa^2}{\pa J_n\pa J_{n+p-1}}-J_{n+p}\frac{\pa}{\pa J_n}-J_{n+p+1}\frac{\pa}{\pa J_{n+p-1}} + (J_{n+p}J_{n+p+1}-1).
\ee  %
We start with ${\cal K}^{(2)}=J_n$ and ${\cal K}^{(3)}=J_nJ_{n+1}-2$.
\el  %
At each step this introduces a new pair of variables $J_{n+p}$ and $J_{n+p+1}$ into the formula.  The next two functions are
\bea  %
{\cal K}^{(4)} &=& J_n J_{n+1} J_{n+2}- (J_n+J_{n+1}+J_{n+2}),\quad\mbox{with}\;\;\; J_{n+3}=J_n, \nn\\
{\cal K}^{(5)} &=& J_n J_{n+1} J_{n+2} J_{n+3}- (J_n J_{n+1}+J_{n+1} J_{n+2}+J_{n+2} J_{n+3}+ J_{n+3} J_n)+2,\quad\mbox{with}\;\;\; J_{n+4}=J_n, \nn
\eea  %
which are just $S_{4,1}$ and $S_{5,1}$ of the list in Example 9.10 of \cite{f11-2}.

\subsubsection{Link with the Poisson Structure}

Now consider the quiver $P_{2m}^{(1)}$ for fixed $p=2m-1$, with periodicity ${\bf L}_{n+p}={\bf L}_n$.  The relation
$$
{\bf M}_n{\bf L}_n={\bf L}_n{\bf M}_{n+1}
$$
leads to $\RC_{n+1}=-\RB_n$, so $\RC_n=-\RB_{n-1}=-\RB_{n+p-1}$, and
\be\label{ml=lm}  %
J_n\RB_n=\RA_n-\RD_{n+1},\quad \RB_{n+p-1}-\RB_{n+1}=J_n(\RD_{n+1}-\RD_n).
\ee  %
Consider the scalar product
\bea  %
&& (0,J_{n+1},-J_{n+2},\dots ,J_{n+p-2},-J_{n+p-1})\cdot (B_n,B_{n+1},\dots ,B_{n+p-1})=\RA_{n+1}-\RD_{n+2}-\RA_{n+2}+\RD_{n+3}+\dots \nn\\
  &&  +\RA_{n+p-2}-\RD_{n+p-1}-\RA_{n+p-1}+\RD_{n+p} = \RA_{n+1}-{\cal K}+\RD_n=\RD_n-\RD_{n+1},  \nn
\eea  %
where we have used $\RA_k+\RD_k={\cal K}$ and $\RD_{n+p}=\RD_n$.  The alternating sign (and the fact that $p=2m-1$) means that only one of the $\cal K$ survives the sum.  Multiplying this scalar product by $J_n$ and using the second of (\ref{ml=lm}), leads to
$$
(0,J_n J_{n+1},-J_n J_{n+2},\dots ,J_n J_{n+p-2},-J_n J_{n+p-1})\cdot (B_n,B_{n+1},\dots ,B_{n+p-1}) = \RB_{n+1}-\RB_{n+p-1},
$$
together with similar expressions obtained by cyclic permutations.  We can form a matrix equation
$$
({\bf P}^{(2)}+{\bf P}^{(0)}){\bf b}=0,\quad \mbox{with}\quad {\bf b}
=(\RB_n,\RB_{n+1},\dots ,\RB_{n+p-1})^T= -\nabla {\cal K},
$$
where
\be\label{p20}
{\bf P}^{(2)}_{ik} =  c^{(2)}_{ik} J_i J_k,\quad {\bf P}^{(0)}_{ik} = c^{(0)}_{ik},
\ee  %
where $c^{(2)}_{ik}$ and $c^{(0)}_{ik}$ are the components of skew-symmetric Toeplitz matrices with top rows given by
$$
(c^{(2)}_{1k}) = (0,1,-1,\cdots ,1,-1),\quad (c^{(0)}_{1k})=(0,1,0,\cdots ,0,-1).
$$
\br  %
Remarkably, these are the Poisson tensors which appear in Lemma \ref{jbra} below, and which are derived from the Poisson bracket $\{x_i,x_k\}$, itself coming from the inversion of the matrix $B$, defining the corresponding quiver.  On the other hand, the quiver has played no role in the manipulations since Equation (\ref{frieze}).
\er  %

We can summarise these results in
\bt \label{isomonodromy}
The function $\cal K$, defined by (\ref{k2}) is the Casimir of the Poisson bracket of Lemma \ref{jbra}:
\be\label{pnabk}  %
({\bf P}^{(2)}+{\bf P}^{(0)})\nabla {\cal K}=0.
\ee  %
\et

\subsection{Poisson Brackets and Liouville Integrability for $P_{2m}^{(1)}$}\label{pn1-even}

The matrix $B$ in the case of $P_{2m}^{(1)}$ is nondegenerate, having the form
\be  \label{btau}  %
B = \tau_N-\tau_N^T, \quad\mbox{with}\quad \tau_N= \sum_{r=1}^{N-1} {\bf E}_{r+1,r} -{\bf E}_{1,N},
\ee  %
where ${\bf E}_{r,s}$ denotes an element of the standard basis for $gl(N)$. This ``skew rotation'' matrix $\tau_N$ already appeared in
Definition \ref{d:skewrotation}.

By Theorem \ref{torusred}, the map is symplectic, and hence there is a nondegenerate Poisson bracket of the form (\ref{logcan}), with
$C=B^{-1}$, up to scaling. We can take
$$
C =\tau_N^T+(\tau_N^T)^3+\dots +(\tau_N^T)^{N-1}= \sum_{s=1}^{\frac{N}{2}}\sum_{r=1}^{N-2s+1} \left(
{\bf E}_{r,r+2s-1} - {\bf E}_{r+2s-1,r}\right),
$$
so that $CB=2\,I$, which gives
\be\label{xbra}
\{ \, x_j , x_k \, \}=\left\{ \begin{array}{ll}
                               \mathrm{sgn}(k-j)\, x_j x_k, \qquad & k-j \quad \mathrm{odd}, \\[2mm]
                               0 & \mathrm{otherwise},
                               \end{array}  \right.
\ee
for $j,k = 1,\ldots , N$, with $\mathrm{sgn}$ denoting the sign function.  We can then directly calculate the Poisson brackets $\{ \, J_i, J_k\, \}$.

\bl \label{jbra}
For even $N=p+1$ and $q=1$, the functions $J_n$ given by (\ref{Jrec}) define a Poisson subalgebra of
codimension one in the algebra (\ref{xbra}), with brackets
\be\label{jbrac}
\{ \, J_i, J_k\, \} = 2 P_{ik},\quad\mbox{where matrix}\quad (P_{ik}) = {\bf P}^{(2)}+{\bf P}^{(0)},
\ee
for $i,k=1,\ldots, p$, and ${\bf P}^{(k)}$ are the matrices defined by (\ref{p20}).
\el
The bracket for the $J_n$ is clearly a sum $\{\, ,\,\}=\{\, ,\,\}_2 + \{\, ,\,\}_0$, corresponding
to the splitting of the Poisson tensor into homogeneous parts ${\bf P}^{(2)}+{\bf P}^{(0)}$, each of which is itself a Poisson bracket.  Hence we have a pair of compatible Poisson brackets $\{\, ,\,\}_k$ to which we can apply the theory of Section \ref{sec:biham}.  Indeed, we have formula (\ref{pnabk}), in which $\cal K$ is given by the repeated action of ${\cal R}^{(p)}$ on ${\cal K}^{(2)}=J_n$ (see (\ref{krecursion}) for $p=2m-1$), so
\be\label{kgen}
\mathcal{K} = \sum_{j=0}^{m-1}(-1)^{m+j+1}\mathcal{I}_j,
\ee
where $\mathcal{I}_j$ is the term of degree $2j+1$ in the variables $J_i$ (compare with (\ref{casimir})).  The homogeneous parts of formula (\ref{pnabk}) can then be written
\be   \label{ladderpn1}  %
{\bf P}^{(0)} \nabla \mathcal{I}_0=0, \;\; {\bf P}^{(0)}\nabla \mathcal{I}_k = {\bf P}^{(2)} \nabla \mathcal{I}_{k-1} ,
\;\;\;\mbox{for}\;\;\; 1\leq k\leq m-1,\;\;\;\mbox{and}\;\;\; {\bf P}^{(2)}\nabla \mathcal{I}_{m-1}=0.
\ee  %
It follows that
$$
\{ \mathcal{I}_j, \mathcal{I}_k \}_0 = 0 = \{ \mathcal{I}_j, \mathcal{I}_k \}_2, \qquad
\mathrm{for} \,\,\, \mathrm{all}\quad j,k.
$$

In summary, we have
\bt[Complete Integrability]
For $N=2m$ and $q=1$ the map $\varphi$ defined by (\ref{jkrec}) has $m$ functionally independent Poisson
commuting integrals, given by the terms of each odd homogeneous degree in the quantity ${\cal K}$, as given by equation (\ref{kgen}).
The map is also maximally super-integrable, having a total of $N-1$ independent first integrals.
\et

\subsubsection{Examples}

I present here some low dimensional examples.  Recall that, in terms of the variables $x_k$, each of these integrals $\mathcal{I}_k$ is a rational function with complicated numerator, but with simple denominator, given by $\prod_{k=1}^{2m}x_k$.

\bex[The Case $N=4$]  {\em   %
Here we have $3$ basic functions $J_1, J_2, J_3$.  With, $m=2$, we have $\mathcal{I}_0=J_1+J_2+J_3$
and $\mathcal{I}_1=J_1J_2J_3$.
}\eex  %

\bex[The Case $N=6$]  {\em   %
Here we have $5$ basic functions $J_1, \dots , J_5$.  With, $m=3$ we have
$$
 \mathcal{I}_0=\sum_{i=1}^{5} J_i , \quad \mathcal{I}_1 = \sum_{i=1}^{5} J_iJ_{i+1}J_{i+2},\quad
 \mathcal{I}_2=\prod_{i=1}^5 J_i,
$$
the indices here being taken modulo $5$.
}\eex  %

\bex[The Case $N=8$]  {\em   %
Here we have $7$ basic functions $J_1, \dots , J_7$.  With $m=4$, we have
$$
 \mathcal{I}_0=\sum_{i=1}^{7} J_i , \quad \mathcal{I}_1 = \sum_{i=1}^{7} J_{i}J_{i+1}(J_{i+2}+J_{i+4}),\quad
 \mathcal{I}_2 = \sum_{i=1}^{7} J_iJ_{i+1}J_{i+2}J_{i+3}J_{i+4}, \quad   \mathcal{I}_3=\prod_{i=1}^7 J_i,
$$
the indices here being taken modulo $7$.
}\eex  %

\subsection{Primitives of the Form $P_{2m+1}^{(1)}$}\label{pn1-odd}

The recurrences (\ref{jkrec}) for $p=2m$ and $q=1$ are given by
\be\label{oddprim}
x_{n+2m+1}\, x_n = x_{n+2m}\, x_{n+1} + 1.
\ee
The formula (\ref{btau}) still holds, but now the matrix $B$ is singular with a one-dimensional kernel,
spanned by the vector ${\bf u}=(1,-1,\dots ,1,-1,1)^T$. In the coordinates
\be\label{oddprimy}
y_j = x_j\, x_{j+1}, \qquad j=1,\ldots , 2m,
\ee
the degenerate form (\ref{omega}) gives rise to a symplectic form (\ref{yform}) in dimension
$2m$, whose coefficients $\hat{b}_{jk}$ are the matrix elements of
$$
\hat{B} = \tau_{2m}-\tau_{2m}^2+\tau_{2m}^3-\dots +\tau_{2m}^{2m-1},
$$
where $\tau_{2m}$ is the $2m\times 2m$ version of $\tau_N$.  The inverse of this
is the skew-symmetric matrix
$$
\hat{B}^{-1}=\tau_{2m}^T+(\tau_{2m}^T)^2+(\tau_{2m}^T)^3+\dots +(\tau_{2m}^T)^{2m-1},
$$
with all components above the diagonal equal to $1$, giving a nondegenerate Poisson bracket for the $y_j$:
\be \label{ybra}
\{ \, y_j , y_k\,\} = \mathrm{sgn}(k-j)\, y_j y_k , \qquad 1\leq j,k\leq 2m.
\ee
In the variables $y_i$, the map (for $m\geq 2$) is given by
\be  \label{ymap}  %
\hat\varphi: \, (y_1,y_2,\ldots , y_{2m-1},y_{2m})
\mapsto (y_2,y_3,\ldots , y_{2m},y_{2m+1}),
\ee
where
$$
y_{2m+1} = y_2y_4\cdots y_{2m}
(y_2y_4\cdots y_{2m}+y_3\cdots y_{2m-1})/(y_1y_3^2\cdots y_{2m-1}^2).
$$
The map is simpler for the case $m=1$, given in Example \ref{ymap-p31} below.

The polynomial $\cal K$ is given by the repeated action of ${\cal R}^{(p)}$ on ${\cal K}^{(3)}=J_nJ_{n+1}-2$ (see (\ref{krecursion}) for $p=2m$), so can be expanded as
\be\label{koddgen}
\mathcal{K} = \sum_{j=0}^{m}(-1)^{m+j}\mathcal{I}_j, \qquad \mathrm{where} \qquad \mathcal{I}_0 = 2,
\ee
and each polynomial $\mathcal{I}_j$ is homogeneous of degree $2j$ in the variables $J_n$.  The non-trivial homogeneous components $\mathcal{I}_1, \ldots , \mathcal{I}_{m}$ provide $m$ first integrals for (\ref{oddprim}) and can be rewritten in terms of the variables $y_j$.

\bex[The primitive $P_3^{(1)}$] \label{ymap-p31} {\em   %
For $m=1$ the map $\varphi$ for the $x_j$ variables  is (\ref{affA2}),
which is associated with $P_3^{(1)}$.  The matrix $B$ has rank two, so in terms of the variables $y_j$ given by
(\ref{oddprimy}) the induced map of the plane is
$$
\hat\varphi: \qquad \left(\begin{array}{c} y_1 \\
                                 y_2 \end{array}\right) \mapsto
                                 \left(\begin{array}{c} y_2 \\
                                 y_2(y_2+1)/y_1
                                 \end{array}\right).
$$
The function
$$
{\cal I}_1= J_1J_2 ={\cal K}+2=\frac{(y_1+y_2)(y_1+y_2+1)}{y_1y_2}
$$
is invariant under $\hat\varphi$, along with the Poisson bracket $\{y_1,y_2\}=y_1y_2$, so the map is integrable.
}\eex  %

\bex[The primitive $P_5^{(1)}$] \label{ymap-p51} {\em   %
For $m=2$ the recurrence (\ref{oddprim}) has the first integrals ${\cal I}_1$ and ${\cal I}_2$ given by the homogeneous terms of degree 2 and 4, respectively, in the expression (\ref{koddgen}):
\be\label{kp51}
{\cal K} = 2-(J_1J_2+J_2J_3+J_3J_4+J_4J_1)+ J_1J_2J_3J_4={\cal I}_0 - {\cal I}_1+{\cal I}_2.
\ee
The variables $y_j$, defined by (\ref{oddprimy}), are endowed with the nondegenerate
Poisson bracket (\ref{ybra}), which is invariant under the map (\ref{ymap}):
$$
\hat\varphi: \quad
(y_1,y_2,y_3,y_4)\mapsto
\Big(y_2,y_3,y_4 , y_2y_4(y_2y_4+y_3)/(y_1y_3^2)\Big).
$$
The terms in the formula (\ref{kp51}) can all be expressed via  the functions
\be \label{wi}
w_i=J_i\, J_{i+1},
\ee
so that ${\cal I}_1=w_1+w_2+w_3+w_4$, ${\cal I}_2=w_1w_3$, and these $w_i$ can be written as functions of $y_j$:
$$
w_1=\frac{(y_1+y_2)(y_2+y_3)}{y_2^2},\quad
w_2=\frac{(y_2+y_3)(y_3+y_4)}{y_3^2}, \quad
w_3=\frac{(y_3+y_4)(y_2y_3+y_1y_3^2+y_2^2y_4)}{y_1y_3^2y_4}.
$$
Under the action of $\hat\varphi$, since the $J_n$ cycle with period 4 under $\varphi$, the $w_i$ transform as
$$
\hat\varphi^* w_1=w_2,\;\; \hat\varphi^*w_2=w_3,\;\; \hat\varphi^*w_3=w_4=\frac{w_1w_3}{w_2},\;\;
\hat\varphi^*w_4=w_1.
$$
The first three $w_i$ form a three-dimensional Poisson subalgebra  of the $y_j$, which
is non-polynomial:
$$
\{w_1,w_2\}=w_1w_2-w_1-w_2,\quad \{w_1,w_3\}=w_2-\frac{w_1w_3}{w_2}, \quad
\{w_2,w_3\}=w_2w_3-w_2-w_3.
$$
The Casimir of this algebra is
$
{\cal I}_1-{\cal I}_2=w_1+w_2+w_3+\frac{w_1w_3}{w_2}-w_1w_3 =2-{\cal K},
$
and $\{ {\cal I}_1,{\cal I}_2\} = \{{\cal I}_1,{\cal K}\} =0$, so the two first
integrals are in involution, as required.
}\eex  %

\section{Linearisable Recurrences from $P_{2m}^{(q)}-P_{2m}^{(m)}+P_{2(m-q)}^{(m-q)}$ quivers}\label{pert}
\setcounter{equation}{0}

In this section we consider the family of recurrences (\ref{comprec}) (case (iii) of Theorem \ref{zeroe}), which come from quivers  of the form
$P_{2m}^{(q)}-P_{2m}^{(m)}+P_{2(m-q)}^{(m-q)}$.  At the beginning of this section I give some general formulae, but to avoid complications I only discuss complete integrability in the context of examples with $q=1$.

\begin{figure}[htb]
\centering
\psfrag{1}{$1$}\psfrag{2}{$2$}\psfrag{3}{$3$}\psfrag{4}{$4$}\psfrag{5}{$5$}\psfrag{6}{$6$}
\subfigure[Example \ref{ex-p4142}.]{
\includegraphics[width=3.5cm]{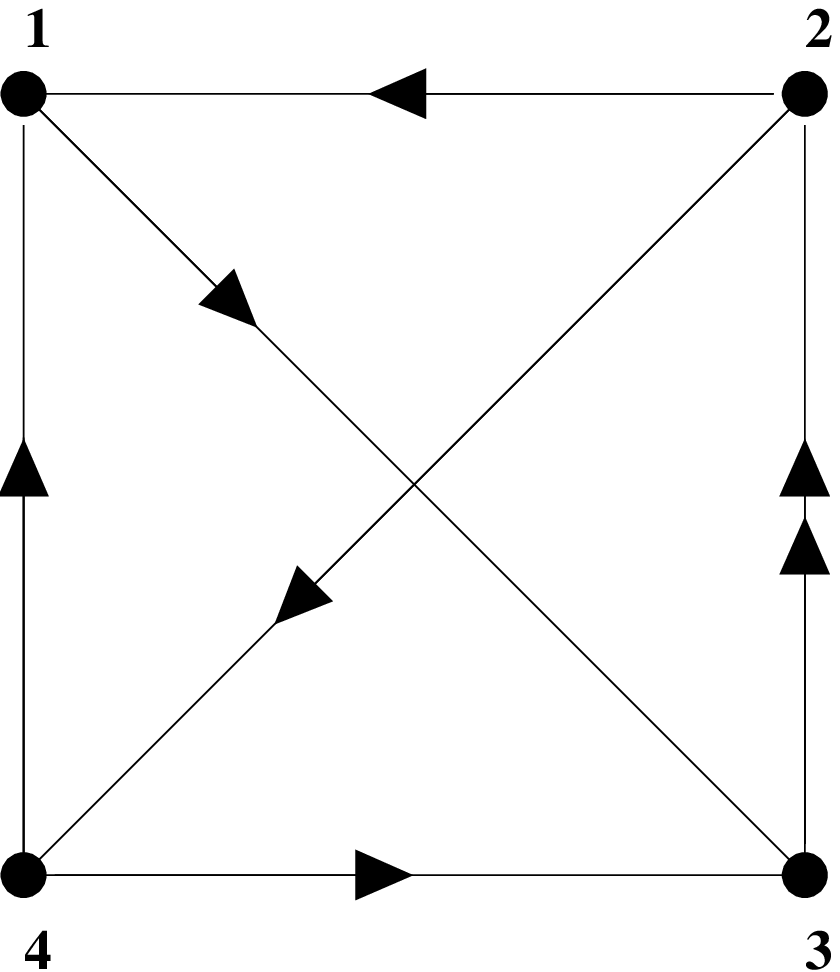}\label{subfig:p4142fig}
} \qquad\qquad
\subfigure[Example \ref{ex-p6163}.]{
\includegraphics[width=4cm]{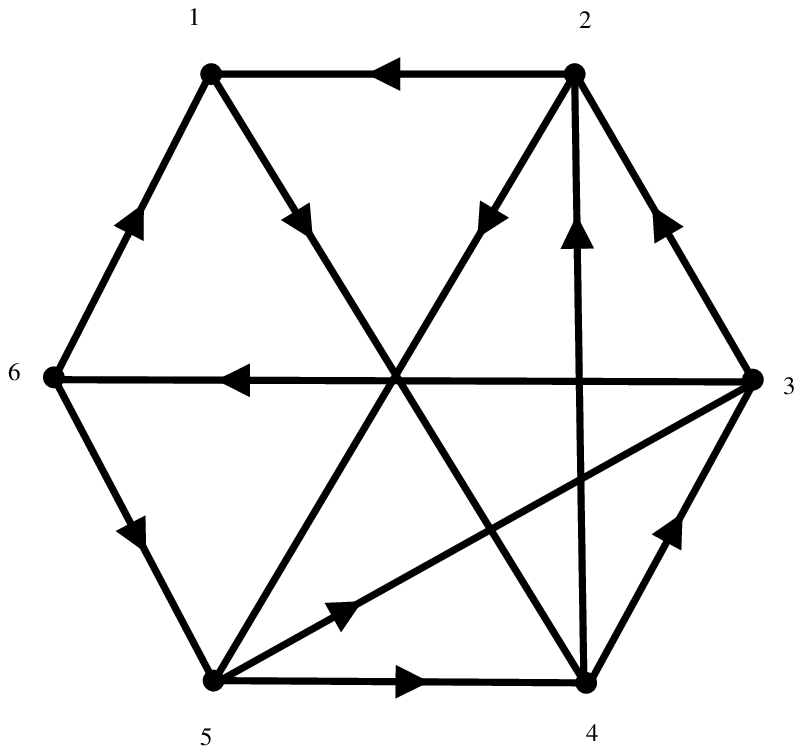}\label{subfig:p6163fig}
} \qquad\qquad
\subfigure[$P_{6}^{(2)}-P_{6}^{(3)}+P_2^{(1)}$]{
\includegraphics[width=4cm]{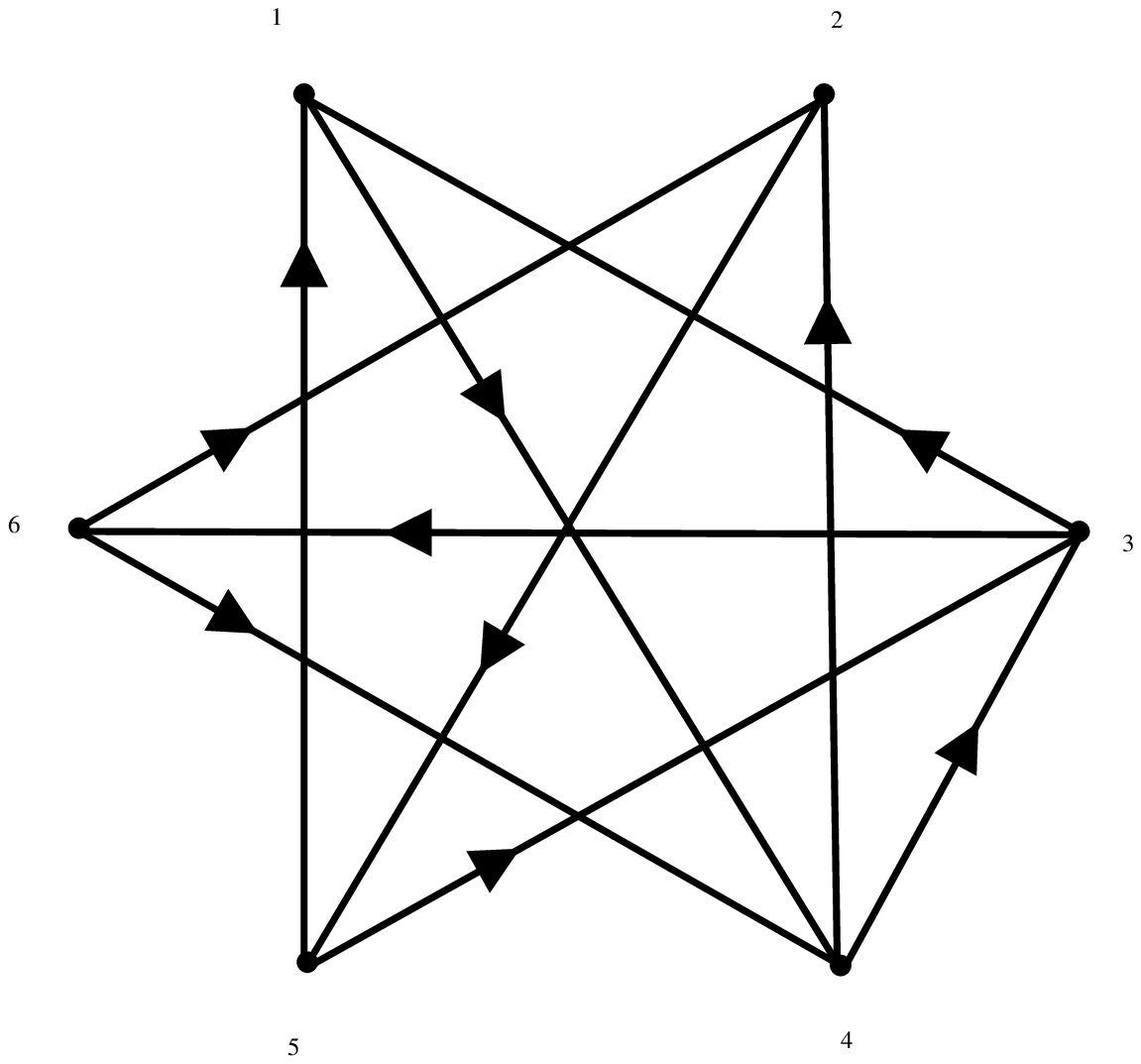}\label{subfig:p6263fig}
} \caption{The first three quivers in this class.}\label{fig:pertquivers}
\end{figure}

The general recurrence can be written as
\be
x_{n+N}\, x_n = x_{n+p}\, x_{n+q} + x_{n+m}, \qquad p+q=N=2m,
\label{pqrec}
\ee
giving (for fixed $m$) a different recurrence for each $q=1,\ldots,m-1$.
The associated matrix $B$ is given by
\be  \label{btau3}  %
B = \tau_{2m}^q-(\tau_{2m}^q)^T-\tau_{2m}^m+\hat\tau_{2(m-q)}^{m-q},
\ee  %
where $\tau_N$ is defined in (\ref{btau}), and $\hat\tau_{2(m-q)}$ denotes
the $N\times N$ matrix obtained by adding $q$ left and right columns and upper and lower rows of zeros
to the $2(m-q)\times 2(m-q)$ matrix $\tau_{2(m-q)}$. The simplest examples of the quivers corresponding to such $B$ are shown in
Figure \ref{fig:pertquivers}.

We can assume that $\mbox{gcd}(m,q)=1$, otherwise the quiver (and the recurrence) decouple.
It then follows that $\mbox{gcd}(p,q)$ can only be $1$ or $2$.  The case $\mbox{gcd}(p,q)=1$ has a very similar structure to that of the
primitives $P_{N}^{(q)}$ for even $N$ and will be the only case discussed in detail here.  The case $\mbox{gcd}(p,q)=2$ has
several new features (see \cite{f14-1}).  The quiver of Figure \ref{subfig:p6263fig} is in this class.

\subsection{Linear Relations with Periodic Coefficients}

By analogy with (\ref{frieze}), the recurrence (\ref{pqrec})  can be written in the form
\be \label{2frieze}
\det \,\Psi_n = \left| \begin{array}{cc} x_n & x_{n+q} \\
                                 x_{n+p} & x_{n+N}
                           \end{array}\right| = x_{n+m}.
\ee
Using Dodgson condensation on $\tilde{\Psi}_n$ (see (\ref{3by3})) now gives
$$
\det\tilde{\Psi}_n =(x_{n+m}x_{n+N+m}-x_{n+m+q}x_{n+m+p})/x_{n+N}=1.
$$
Then condensing the appropriate  $4\times 4$ matrix $\Delta_n$ in terms of
$3\times 3$ minors yields
$$
\det \Delta_n =  \left| \begin{array}{cccc}x_n & x_{n+q} & x_{n+2q} &  x_{n+3q}\\
                               x_{n+p} & x_{n+N} & x_{n+N+q} & x_{n+N+2q} \\
                                 x_{n+2p} & x_{n+N+p} & x_{n+2N} & x_{n+2N+q} \\
                                    x_{n+3p} & x_{n+N+2p} & x_{n+2N+p} & x_{n+3N}
                                \end{array}\right|=0.
$$
The left and right kernels of the singular matrix  $\Delta_n$ yield linear relations between the $x_n$.
\bl \label{PQ}
The iterates of the recurrence (\ref{pqrec}) satisfy the linear relations
\bea
&& x_{n+3q}-J_{n+m}\,x_{n+2q}+ J_{n}\, x_{n+q} - x_n = 0,  \label{Prec}  \\
&& x_{n+3p}-K_{n+m}\,x_{n+2p}+ K_{n}\, x_{n+p} - x_n = 0,  \label{Qrec}
\eea
whose coefficients are periodic functions of period $p,q$ respectively:
$$
J_{n+p}=J_n, \qquad K_{n+q}=K_n, \qquad  for\,\,  all \, \, n.
$$
\el

\br\label{penta}  %
The four-term linear relations (\ref{Prec}) and (\ref{Qrec}),
together with $\det\tilde{\Psi}_n =1$, should be compared with those of the
pentagram map \cite{10-2}, but there the coefficients of the second and third terms are independent.
\er  %

\subsection{Monodromy and Linear Relations with Constant Coefficients}
\label{3x3monodromy}

This subsection follows closely the discussion of subsection \ref{2x2monodromy} for the case of primitives.

The relations (\ref{Prec}) and (\ref{Qrec}) imply that the matrix $\tilde\Psi_{n}$ satisfies
\bea
&& \tilde\Psi_{n+q}=\tilde\Psi_n\, {\bf L}_n, \qquad
                  {\bf L}_n=\left(\begin{array}{ccc}
                                 0 & 0 & 1 \\
                                 1 & 0 & -J_n \\
                                 0  & 1 & J_{n+m}
                                 \end{array}\right), \quad\mbox{with}\;\; {\bf L}_{n+p}={\bf L}_n, \label{psi3rel1}  \\
&& \tilde\Psi_{n+p}= \hat{\bf L}_n \,\tilde\Psi_n, \qquad
                         \hat{\bf L}_n=\left(\begin{array}{ccc}
                                                   0 &  1 & 0 \\
                                                   0 & 0 & 1 \\
                                                   1  & -K_n & K_{n+m}
                                                   \end{array}\right), \quad\mbox{with}\;\; \hat{\bf L}_{n+q}=\hat{\bf L}_n.  \label{psi3rel2}
\eea
Taking appropriate ordered products then leads to the monodromy matrices
\bea
&&  {\bf M}_n:=  {\bf L}_{n}{\bf L}_{n+q}\ldots{\bf L}_{n+(p-1)q}=\tilde\Psi_n^{-1}\,\tilde \Psi_{n+pq},  \label{mdy31}  \\
&&  \hat{{\bf M}}_n:=  \hat{\bf L}_{n+(q-1)p}\ldots\hat{\bf L}_{n+p}\hat{\bf L}_n=\tilde\Psi_{n+pq}\, \tilde\Psi_n^{-1}.  \label{mdy32}
\eea
From the cyclic property of the trace it follows that
\be\label{k32}
\mathcal{K}_n:= \mathrm{tr}\, {\bf M}_n = \mathrm{tr}\, \hat{\bf M}_n.
\ee
Once again, the periodicity of ${\bf L}_n$, implies that $\mathcal{K}_{n+p}=\mathcal{K}_n$, and $\mathcal{K}_{n+q}=\mathcal{K}_n$, so,
if the periods $p$ and $q$ are coprime, $\mathcal{K}_n=\mathcal{K}=$~constant, for all $n$,
hence $\mathcal{K}$ is a first integral for the map $\varphi$ corresponding to (\ref{pqrec}).

In \cite{f14-1} we prove the following:
\bt\label{Klin3}  %
When $\mbox{gcd}(p,q)=1$, the iterates of the nonlinear recurrence (\ref{pqrec}) satisfy the linear relation
\be\label{krec3}
x_{n+3pq}- \mathcal{K}\,x_{n+2pq}+\mathcal{K}\,x_{n+pq}-x_n=0,
\ee
where  $\mathcal{K}$ is the first integral defined by (\ref{k32}).
\et  %

\br  \label{gcd1} %
In particular, when $q=1$ the coefficient $K_n$ has period 1, so $K_{n+1}=K_n={\cal K}$
for all $n$, and the recurrence (\ref{Qrec}) is just the
{\it constant coefficient, linear difference equation}
\be\label{q1krec}
x_{n+3N-3} -{\cal K}\, x_{n+2N-2}+{\cal K}\, x_{n+N-1}-x_n =0.
\ee
This is the case we now consider.
\er  %

\subsection{The Case $q=1$ and Examples}

The  discussion of this case is almost identical to that for the
primitives $P_N^{(1)}$ with $N$ even. The matrix $B$ is nondegenerate, so we get an
invariant Poisson bracket of the form (\ref{logcan}),
which is specified uniquely (up to scale) by the Toeplitz matrix $C=B^{-1}$.
However, in this case a general closed-form expression for the entries of $C$, analogous
to the formula (\ref{xbra}) for the Poisson bracket of the even primitives, is not available to us at present.

The functions $J_n$ are of period $p$, while $K_n={\cal K}= \mbox{constant}$, with
the first integral $\mathcal{K}$, defined by (\ref{k32}), a cyclically symmetric polynomial
in the $J_n$, $n=1,\ldots, p$.

From the log-canonical bracket $\{x_i,x_k\}$ it is possible to derive the brackets $\{J_i,J_k\}$.
In general (excluding $m=2$), this Poisson bracket consists of three homogeneous parts (of degrees $0, 1$ and $2$), which
(unlike the bracket (\ref{jbrac}) for the primitive $P_{2m}^{(1)}$) does not have an obvious splitting into a bi-Hamiltonian pair.
Moreover, while every cyclically symmetric polynomial function of the $J_n$ is a first integral
of the map $\varphi$, we do not yet have an algorithm for selecting an involutive set of them.

\bex[The quiver $P_{4}^{(1)}-P_{4}^{(2)}+P_{2}^{(1)}$] \label{ex-p4142} {\em   %
This example was first studied in an ad hoc way in \cite{07-2}.
The appropriate matrix (\ref{btau3}),
which corresponds to the quiver in Figure \ref{subfig:p4142fig}, is obtained by
setting $c=1$ in (\ref{s4gen}).  The explicit form of the map is
\be  \label{p4142map}  %
\varphi : \quad (x_1,x_2,x_3,x_4) \mapsto  (x_2,x_3,x_4,x_5), \quad  x_5=  \frac{x_2x_4+x_3}{x_1}.
\ee  %
The invariant Poisson bracket (\ref{logcan}) for this map is given by the
Toeplitz matrix $C=2B^{-1}$, with top row  $(c_{1,j})=(0,1,1,2)$.  The period 3 functions $J_i$ take the form
$$
J_1=\frac{x_3(x_2+x_1x_3)+x_4(x_1^2+x_2^2)}{x_1x_2x_4},\quad
J_2=\frac{x_1x_4+x_2^2+x_3^2}{x_2x_3},\quad
J_3=\frac{x_3(x_2+x_1x_3)+x_4(x_1x_4+x_2^2)}{x_1x_3x_4}.
$$
The Poisson brackets between these three functions follow by the cyclic property from
\be\label{N4jbr}
\{J_1,J_2\}=J_1J_2-2J_3.
\ee
This example is exceptional, in that the bracket for the $J_i$ is the sum of only {\em two} homogeneous terms:
$$
{\bf P}={\bf P}^{(2)}+{\bf P}^{(1)},\quad\mbox{where}\;\;\;
     {\bf P}^{(2)}=\left(\begin{array}{ccc}
                   0 & J_1J_2 & -J_1J_3 \\
                  -J_1J_2 & 0 & J_2J_3 \\
                   J_1J_3 & -J_2J_3 & 0
                      \end{array} \right), \quad
                       {\bf P}^{(1)}=\left(\begin{array}{ccc}
                   0 & -2J_3 & 2J_2 \\
                   2J_3 & 0 & -2J_1 \\
                   -2J_2 & 2J_1 & 0
                      \end{array} \right).
$$
Each of the tensors specified by ${\bf P}^{(1)}$ and ${\bf P}^{(2)}$
satisfies the Jacobi identity, so (since their sum is a Poisson tensor)
they define a compatible pair of Poisson brackets.

The first integrals
$$
\CH_1=J_1^2+J_2^2+J_3^2, \qquad \CH_2=J_1J_2J_3
$$
satisfy the bi-Hamiltonian ladder %relations
$$
 {\bf P}_1\nabla \CH_1=0,\qquad  {\bf P}_2\nabla \CH_1={\bf P}_1\nabla \CH_2,
\qquad {\bf P}_2\nabla \CH_2=0,
$$
so they commute with respect to the bracket defined by (\ref{N4jbr}).
The quantity
\be\label{k31eq}
{\cal K}=3-\CH_1+\CH_2
\ee
provides the Casimir of this bracket.
}\eex  %

\bex[The quiver $P_{6}^{(1)}-P_{6}^{(3)}+P_{4}^{(2)}$] \label{ex-p6163} {\em   %

Mutation of the quiver in Figure \ref{subfig:p6163fig} gives the map
$$
\varphi : \quad (x_1,x_2,x_3,x_4,x_5,x_6) \mapsto  (x_2,x_3,x_4,x_5,x_6,x_7), \quad
  x_7=  \frac{x_2x_6+x_4}{x_1}.
$$
The invariant Poisson bracket for this map is given by (\ref{logcan}),
with the coefficients specified by the Toeplitz matrix $C=B^{-1}$ with top row
$(c_{1,j})=(0,0,1,0,1,1)$.
The functions $J_i$, which cycle with period 5 under the action of $\varphi$, take the form
\bea  %
&& J_1=\frac{x_1x_5+x_2x_6+x_3x_4}{x_2x_5},\quad
J_2=\frac{x_4(x_3+x_1x_5)+(x_1+x_3)x_2x_6}{x_1x_3x_6},\quad
J_3=\frac{x_3(x_2+x_4)+x_1x_5)}{x_2x_4},  \nn\\
&& J_4=\frac{x_4(x_3+x_5)+x_2x_6)}{x_3x_5},\quad
J_5=\frac{x_4(x_3+x_1x_5)+(x_1x_5+x_2x_3)x_6}{x_1x_4x_6}.  \nn
\eea  %
The Poisson bracket between these five functions follow by the cyclic property
from
$$
\{J_1,J_2\}=-J_1J_2-J_4+1,\quad \{J_1,J_3\}=2J_1J_3.
$$
This Poisson bracket is the sum of three homogeneous terms,
$$
{\bf P}={\bf P}^{(2)}+{\bf P}^{(1)}+{\bf P}^{(0)}, \quad\mbox{with}\;\;\; {\bf P}^{(2)}_{ik}=c^{(2)}_{ik}J_iJ_k,\;\;\; {\bf P}^{(1)}_{ik}=c^{(1)}_{ik}J_{k+2},\;\;\;
   {\bf P}^{(0)}_{ik}=c^{(0)}_{ik},
$$
where $c^{(\ell)}_{ik}$ are the Toeplitz matrices with top rows given by
$$
(c^{(2)}_{1,k})=(0,-1,2,-2,1),\quad (c^{(1)}_{1,k}) =(0,-1,0,0,1),
\quad (c^{(0)}_{1,k})=(0,1,0,0,-1).
$$
Both ${\bf P}^{(0)}$ and ${\bf P}^{(2)}$ satisfy the Jacobi identity,
but ${\bf P}^{(1)}$ does not, so we cannot think of this sum as some sort of Poisson compatibility.

The Casimir for the 5-dimensional Poisson algebra generated by the $J_i$ is the trace of the monodromy matrix,
as in (\ref{k32}).  It can be written as the sum
\be\label{k51}
{\cal K}=\CH_3+\CH_2-\CH_1,
\ee
where each of the components is a first integral:
$$
\CH_1=\sum_{i=1}^5(J_i-J_iJ_{i+1}),\quad
\CH_2=\sum_{i=1}^5(J_iJ_{i+1}J_{i+2}-J_iJ_{i+1}^2J_{i+2}),\quad
\CH_3=\prod_{i=1}^5J_i.
$$
We find that $\{ \CH_i ,\CH_j \} =0$ for all $i,j$, so the 6-dimensional
Poisson map $\varphi$ has the correct number of first integrals in involution.
}\eex  %

\bex[The quiver $P_{8}^{(1)}-P_{8}^{(4)}+P_{6}^{(3)}$] \label{ex-p8184} {\em   %

The map obtained from mutation of this quiver is
$$
\varphi : \quad (x_1,x_2,x_3,x_4,x_5,x_6,x_7,x_8)
\mapsto  (x_2,x_3,x_4,x_5,x_6,x_7,x_8,x_9), \quad x_9=  \frac{x_2x_8+x_5}{x_1}.
$$
The corresponding non-singular matrix $B$ defines an invariant Poisson bracket (\ref{logcan})
with matrix $C=B^{-1}$. The top row of the Toeplitz matrix $C$ is
$(c_{1,j})=(0,1,0,0,1,1,0,1)$.
The functions $J_i$ with period 7 can be determined from $J_1$, which takes the form
$$
 J_1=\frac{x_1x_6+x_2x_7+x_3x_5}{x_2x_6}.
$$
The remaining six functions are obtained by applying $\varphi^*J_i=J_{i+1}$, with $(\varphi^*)^7J_i = J_i$.
The Poisson bracket relations between these functions follow by the cyclic property from
$$
\{J_1,J_2\}=2J_1J_2-J_5,\quad \{J_1,J_3\}=-J_1J_3+1,\quad
\{J_1,J_4\}=-J_1J_4.
$$
Again, this Poisson bracket is the sum of three homogeneous terms,
${\bf P}={\bf P}^{(2)}+{\bf P}^{(1)}+{\bf P}^{(0)}$,
where ${\bf P}^{(0)}$ and ${\bf P}^{(2)}$ satisfy the Jacobi identity,
but ${\bf P}^{(1)}$ does not.

The Casimir of the Poisson subalgebra generated by the $J_i$
is again $\cal K$ (the trace of the monodromy matrix), but in this case
it is not clear how to split the Casimir into four pieces that Poisson commute,
as required for Liouville integrability.  Of course, there are still
seven functionally independent invariant functions (built from cyclically symmetric combinations of the $J_i$)
and we expect that four commuting functions exist.
}\eex  %

\section{Gale-Robinson Recurrences} \label{grmaps}

\setcounter{equation}{0}

We now consider the final class of recurrence (\ref{somosN}),
\be \label{GR}
x_{n+N}\, x_n = x_{n+N-p}\, x_{n+p} + x_{n+N-q}\, x_{n+q},
\ee
listed in Theorem \ref{zeroe} (case (iv)).  These are known as three-term Gale-Robinson recurrences \cite{91-17} and include Somos-$4$ (\ref{somos4}), Somos-$5$ (\ref{somos5}) and the reductions of Somos-$6$ (see Equations (\ref{gr1}), (\ref{gr2}), and (\ref{gr3})) as special cases.  We can no longer expect to linearise these, since their solutions are given in terms of elliptic functions \cite{05-3,07-6,10-3}.

The  relations (\ref{GR}) all arise by reduction of the Hirota-Miwa (discrete KP) equation (see \cite{97-6}), which is
the bilinear partial difference equation
\be\label{dkp}
T_{1}\, T_{-1}=T_{2}\, T_{-2}+T_{3}\, T_{-3}.
\ee
In the above, $T=T(n_1,n_2,n_3)$ is a function of three independent variables, and shifts are denoted by $T_{\pm j} = T|_{n_j\to n_j\pm 1}$.
If we set
\be\label{redntau}
T(n_1,n_2,n_3)=\exp \left(\sum_{i,j}S_{ij}n_in_j\right)\, \uptau (n),
\ee
where $S=(S_{ij})$ is a symmetric matrix and $n=n_0+\updelta_1n_1+\updelta_2n_2+\updelta_3n_3$, then  $\uptau(n)$
satisfies the ordinary difference equation
$$
\uptau (n+\updelta_1)\uptau(n-\updelta_1)=\alpha \, \uptau(n+\updelta_2)\uptau(n-\updelta_2) +\beta\,  \uptau(n+\updelta_3)\uptau(n-\updelta_3)
$$
where $\alpha = \exp(2(S_{22}-S_{11}))$, $\beta =\exp(2(S_{33}-S_{11}))$. Upon taking
$x_n=\uptau (n-\updelta_1)$ with $\updelta_1=\frac{1}{2}N$, $\updelta_2= \frac{1}{2}(N-2p)$,
$\updelta_3= \frac{1}{2}(N-2q)$, this becomes (\ref{GR}) (with some additional parameters $\alpha, \beta$, which I have dropped from this presentation).

\br[Octahedron Recurrence]  %
In the combinatorics literature, equation (\ref{dkp}) is referred to as the octahedron recurrence,
which has the Laurent property (shown in \cite{02-2}).  The Laurent property for three-term Somos
(or Gale-Robinson) recurrences of the form (\ref{GR}) then follows by the reduction (\ref{redntau}).
\er  %

The Hirota-Miwa equation (\ref{dkp}) has a scalar Lax pair, being the compatibility condition for the linear system given by
\be  \label{kplax} %
 T_{-1,3}\, \psi_{1,2}  +  T\, \psi_{2,3}  =  T_{2,3}\, \psi , \qquad
       T\, \psi_{-1,2}  +  T_{-1,3}\, \psi_{2,-3}  =  T_{-1,2}\, \psi ,
\ee  %
where the scalar function $\psi = \psi (n_1,n_2,n_3)$, with the same notation for shifts as before.
Using the latter, one can use the reduction (\ref{redntau})
to obtain Lax pairs for all of the Gale-Robinson recurrences (\ref{GR}), which leads directly
to spectral curves whose coefficients are conserved quantities.

The process becomes quite involved for large $N$, but for $N=4$ and $N=5$ we can derive Lax pairs for Somos-$4$ and $5$ (in terms of the the reduced variables $y_j$ for the (degenerate) pre-symplectic form (\ref{omega})).

\bex[Somos-$4$ Lax pair]  \label{s4lax}  {\em  %
The Lax pair for the map (\ref{s4map}) takes the form
\be \label{laxs4}
{\bf L} \, {\mathbf w} = \xi {\mathbf w}, \qquad
\tilde{{\mathbf w}} ={\bf M} \, {\mathbf w} ,
\ee
where the tilde denotes the shift $n\to n+1$. The matrices ${\bf L}={\bf L}(\zeta )$,
${\bf M}={\bf M}(\zeta )$ are functions of $y_j$ and the spectral parameter $\zeta$,  given by
$$
{\bf L} = \left(\begin{array}{cc}
                  -\,\frac{y_1+1}{y_1y_2} \, \zeta & - y_1 \zeta + \frac{ y_1+1}{y_1y_2}  \\
                   \frac{1}{y_1}\, \zeta^2 -\zeta & \left(-y_1y_2-\frac{1}{y_1}\right)\, \zeta +1
                  \end{array} \right),
\qquad
{\bf M} = \left(\begin{array}{cc}
                   0 & 1 \\
                   -\frac{1}{y_1y_2}\, \zeta & \frac{1}{y_1y_2}
                   \end{array} \right).
$$
The discrete Lax equation $\tilde{\bf L} {\bf M}={\bf M} {\bf L}$ holds if and only if
the map (\ref{s4map}) does.  The spectral curve is
$$
\mathrm{det}\, ({\bf L}(\zeta ) - \xi \, {\mathbf 1}) = \xi^2+(H_1\, \zeta -1)\xi+ \zeta^3+\zeta^2 =0,
$$
in which the coefficient of $\zeta \xi$ is the first integral
\be\label{s4ham}
H_1=y_1y_2+\frac{1}{y_1}+\frac{1}{y_2}+\frac{1}{y_1y_2}.
\ee
The level sets of $H_1$ are biquadratic curves of genus one in the $(y_1,y_2)$ plane, and the map is
a particular instance of the QRT family \cite{88-5}.
}\eex  %

\bex[Somos-$5$ Lax pair]  \label{s5lax}  {\em  %
The matrix $B$ for the recurrence (\ref{somos5}) has rank $2$, so on symplectic leaves, with coordinates $(y_1,y_2)$, the form (\ref{omega}) reduces to (\ref{s4canon}) (see \cite{f11-3}) and the map takes the form
\be\label{s5map}
\hat\varphi : (y_1\, , \, y_2) \mapsto \Big( y_2\, ,\, (y_2 +1)/(y_1y_2) \Big) .
\ee
This arises as the compatibility condition $\tilde{\bf L} {\bf M}={\bf M} {\bf L}$
for a linear system of the form (\ref{laxs4}), where
\be\label{s5lm}
{\bf L} ={\bf C}_0 +{\bf C}_1\, \zeta + {\bf C}_2 \, \zeta^2, \quad
 {\bf M} ={\bf C}_0 + \left(\begin{array}{cc} 0 & 0 \\
                                    -y_1 & 0 \end{array}
                                    \right)\, \zeta,
\ee
with
$$
{\bf C}_0=\left(\begin{array}{cc} 0 & 1 \\
                       0 & 1
                       \end{array} \right) , \quad
                       {\bf C}_1=\left(\begin{array}{cc}
                                         -y_1 & -\left(y_2+\frac{1}{y_1}\right) \\
                                          -y_1 & -\left(y_2+\frac{1}{y_1} +\frac{1}{y_2}+\frac{1}{y_1y_2}\right)
                                          \end{array} \right) , \quad
                                          {\bf C}_2= \left(\begin{array}{cc}
                                                         1 & 0 \\
                                                         1+\frac{y_1+1}{y_2} & 1
                                                         \end{array} \right).
$$
The coefficient of $\zeta \xi$ in the equation for the spectral curve, that  is
$$
\mathrm{det}\, ({\bf L}(\zeta ) - \xi \, {\mathbf 1}) = \xi^2-(2\zeta^2-J\, \zeta +1)\xi+ \zeta^4+\zeta^3 =0,
$$
gives a first integral whose level sets are cubic (also biquadratic) curves of genus one, that is
\be\label{s5j}
J=y_1+y_2+\frac{1}{y_1}+ \frac{1}{y_2} + \frac{1}{y_1y_2}.
\ee
}\eex  %

\section{Conclusions}

One of the main thrusts of this work has been to understand how to build special types of quiver so that a particular sequence of cluster mutations generated a recurrence.  The relation to cluster algebras guarantees that this recurrence will have the Laurent property.  Given such a recurrence we can consider the associated map and ask about its properties.  The relationship with mutation periodic quivers implies that such maps preserve a pre-symplectic form, leading us to Poisson maps.  For the period $1$ case we classified the cases with zero algebraic entropy and analysed them in the framework of Liouville integrability. In \cite{f11-2} it was shown how to include parameters in the coefficients of our recurrences, using the theory of ``ice quivers'', but for simplicity we have not included this here.  A relationship between period 1 primitive recurrences and Pell's equation was also discussed.  In \cite{f11-1} it was shown that the map associated with the primitive $P_{2m}^{(1)}$ (see Section \ref{pn1-even}) can be seen as B\"acklund transformations for Liouville's equation (a well known partial differential equation, which arises in differential geometry and many other contexts).  In \cite{f14-1} there are many details which I omitted here.  In particular, there are numerous new and interesting examples of integrable maps.

There are many things which are still not understood.  We only have a partial classification of period $2$ quivers and only examples of ones with higher periods.  Almost nothing is known about the integrability of maps associated with quivers of period $2$ and above.  Even in the period $1$ case there are open questions about the Liouville integrability of cases (iii) and (iv) of Theorem \ref{zeroe}.

Recently, with the introduction of ``Laurent phenomenon algebras'' \cite{12-2} it has been possible to include such recurrences as (\ref{s6-sequ}) (Somos-$6$), with more than two terms on the ``right hand side''.  Some of the results of \cite{f11-2} are generalised to this case in \cite{13-1}.

In \cite{f11-2} we showed an intriguing connection between our results and those of {\em quiver gauge theories} and {\em $D-$branes}.  The quivers arising in supersymmetric quiver gauge theories often have periodicity properties.  Indeed, the Somos-$5$ quiver (see Figure~\ref{fig:somos5quiver}) also appears in \cite{07-3} (with a relabelling) in the context of a $dP_2$ brane tiling.  The combinatorial rule for Seiberg-dualising a quiver coincides with the rule for Fomin-Zelevinsky quiver mutation (Definition~\ref{d:mutate}) (see \cite{09-6} for a discussion of this relationship).
Recently, the authors of \cite{13-2} have taken our decomposition (from Theorem \ref{p1-theorem}) of a Gale-Robinson quiver to form an infinite, doubly periodic quiver on the plane, whose dual is a brane tiling.

The existence of these diverse links to many parts of mathematics and physics is one of the remarkable features of cluster algebras, as can be seen elsewhere in this volume.

\subsubsection*{Acknowledgments:}

Most of the work reported here was carried out in collaboration with Robert Marsh and Andy Hone.  The research with Robert Marsh was completely unexpected: it arose out of some questions I had about a short lecture course he gave on cluster algebras and opened up a whole new area of research for me.  The research with Andy Hone began during the Programme on Discrete Integrable Systems, Newton Institute, Cambridge, 2009.

%\bibliography{apf}

\begin{thebibliography}{10}

\bibitem{13-1}
J.~Alman, C.~Cuenca, and J.~Huang.
\newblock {Laurent} phenomenon sequences.
\newblock 2013.
\newblock arXiv:1309.0751 [math.CO].

\bibitem{fb90-4}
M.~Antonowicz and A.P. Fordy.
\newblock Hamiltonian structures of nonlinear evolution equations.
\newblock In A.P. Fordy, editor, {\em Soliton Theory : A Survey of Results},
  pages 273--312. MUP, Manchester, 1990.

\bibitem{78-3}
V.I. Arnol'd.
\newblock {\em Mathematical Methods of Classical Mechanics}.
\newblock Springer-Verlag, Berlin, 1978.

\bibitem{99-12}
M.~Bellon and C-M. Viallet.
\newblock Algebraic entropy.
\newblock {\em Comm.Math.Phys.}, 204:425--37, 1999.

\bibitem{04-4}
A.V. Bolsinov and A.T. Fomenko.
\newblock {\em Integrable Hamiltonian Systems: Geometry, Topology,
  Classification.}
\newblock Chapman and Hall/CRC, Boca Raton, 2004.

\bibitem{1866-1}
C.~L. Dodgson.
\newblock Condensation of determinants.
\newblock {\em Proc. R. Soc. Lond.}, 15:150--55, 1866.

\bibitem{93-8}
I.Ya. Dorfman.
\newblock {\em Dirac Structures and Integrability of Nonlinear Evolution
  Equations}.
\newblock Wiley, Chichester, 1993.

\bibitem{02-3}
S.~Fomin and A.~Zelevinsky.
\newblock Cluster algebras {I}: Foundations.
\newblock {\em J. Amer Math Soc}, 15:497--529, 2002.

\bibitem{02-2}
S.~Fomin and A.~Zelevinsky.
\newblock The {Laurent} phenomenon.
\newblock {\em Advances in Applied Mathematics}, 28:119--144, 2002.

\bibitem{f11-1}
A.P. Fordy.
\newblock Mutation-periodic quivers, integrable maps and associated {Poisson}
  algebras.
\newblock {\em Phil. Trans. R. Soc. A}, 369:1264--79, 2011.
\newblock arXiv:1003.3952v2 [nlin.SI].

\bibitem{f11-3}
A.P. Fordy and A.N.W. Hone.
\newblock Symplectic maps from cluster algebras.
\newblock {\em SIGMA}, 7:091, 12 pages, 2011.
\newblock arXiv:1105.2985v2 [nlin.SI].

\bibitem{f14-1}
A.P. Fordy and A.N.W. Hone.
\newblock Discrete integrable systems and {Poisson} algebras from cluster maps.
\newblock {\em Comm Math Phys}, 325:527--584, 2014.
\newblock arXiv:1207.6072v2 [nlin.SI].

\bibitem{f11-2}
A.P. Fordy and R.J. Marsh.
\newblock Cluster mutation-periodic quivers and associated {Laurent} sequences.
\newblock {\em J Algebr Comb}, 34:19--66, 2011.
\newblock arXiv:0904.0200v5 [math.CO].

\bibitem{91-17}
D.~Gale.
\newblock The strange and surprising saga of the {Somos} sequences.
\newblock {\em Math. Intelligencer}, 13:40--2, 1991.

\bibitem{03-5}
M~Gekhtman, M~Shapiro, and A~Vainshtein.
\newblock Cluster algebras and {Poisson} geometry.
\newblock {\em Moscow Math.J}, 3:899--934, 2003.

\bibitem{05-6}
M.~Gekhtman, M.~Shapiro, and A.~Vainshtein.
\newblock Cluster algebras and {Weil–-Petersson} forms.
\newblock {\em Duke Math. J.}, 127:291--311, 2005.

\bibitem{79-5}
I.M. Gelfand and I.Ya. Dorfman.
\newblock {Hamiltonian} operators and algebraic structures related to them.
\newblock {\em Func.Anal \& Apps.}, 13:13--30, 1979.

\bibitem{05-3}
A.N.W. Hone.
\newblock Elliptic curves and quadratic recurrence sequences.
\newblock {\em Bull. London Math. Soc.}, 37:161–71, 2005.

\bibitem{07-2}
A.N.W. Hone.
\newblock Laurent polynomials and super-integrable maps.
\newblock {\em SIGMA}, 3:022, 18 pages, 2007.

\bibitem{07-6}
A.N.W. Hone.
\newblock Sigma function solution of the initial value problem for {Somos} 5
  sequences.
\newblock {\em Trans. Amer. Math. Soc}, 359:5019--34, 2007.

\bibitem{10-3}
A.N.W. Hone.
\newblock Analytic solutions and integrability for bilinear recurrences of
  order six.
\newblock {\em Applicable Analysis}, 89:473 — 492, 2010.

\bibitem{13-2}
I.~Jeong, G.~Musiker, and S.~Zhang.
\newblock {Gale--Robinson} sequences and brane tilings.
\newblock {\em DMTCS proc. AS}, pages 737--48, 2013.
\newblock http://www.liafa.jussieu.fr/fpsac13/pdfAbstracts/dmAS0169.pdf.

\bibitem{keller}
B.~Keller.
\newblock {Quiver mutation in Java}.
\newblock {\em http://www.math.jussieu.fr/$\sim$keller/quivermutation/}.

\bibitem{12-2}
T.~Lam and P.~Pylyavskyy.
\newblock {Laurent} phenomenon algebras.
\newblock 2012.
\newblock arXiv:1206.2611 [math.RT].

\bibitem{87-13}
S.~Maeda.
\newblock Completely integrable symplectic mapping.
\newblock {\em Proc. Japan Acad.}, 63:198--200, 1987.

\bibitem{78-4}
F.~Magri.
\newblock A simple model of the integrable {Hamiltonian} equation.
\newblock {\em J.Math.Phys}, 19:1156--1162, 1978.

\bibitem{11-2}
T.~Nakanishi.
\newblock Periodicities in cluster algebras and dilogarithm identities.
\newblock In A.~Skowronski and K.~Yamagata, editors, {\em Representations of
  Algebras and Related Topics}, EMS Series of Congress Reports, pages 407--443.
  European Mathematical Society, Zurich, 2011.

\bibitem{08-6}
A.~Nobe.
\newblock Ultradiscrete qrt maps and tropical elliptic curves.
\newblock {\em J. Phys. A: Math. Theor.}, 41:125205, 2008.

\bibitem{86-1}
P.J. Olver.
\newblock {\em Application of Lie Groups to Differential Equations}.
\newblock Springer-Verlag, Berlin, 1986.

\bibitem{07-3}
T.~Oota and Y.~Yasui.
\newblock New example of infinite family of quiver gauge theories.
\newblock {\em Nucl. Phys. B}, 762:377--91, 2007.

\bibitem{10-2}
V.~Ovsienko, R.~Schwartz, and S.~Tabachnikov.
\newblock The pentagram map: a discrete integrable system.
\newblock {\em Commun. Math. Phys.}, 299:409--446, 2010.

\bibitem{88-5}
G.R.W. Quispel, J.A.G. Roberts, and C.J. Thompson.
\newblock Integrable mappings and soliton equations.
\newblock {\em Phys. Letts. A.}, 126:419--421, 1988.

\bibitem{sloane}
N.J.A. Sloane.
\newblock {The On-Line Encyclopedia of Integer Sequences}.
\newblock {\em http://oeis.org/}.

\bibitem{91-4}
A.P. Veselov.
\newblock Integrable maps.
\newblock {\em Russ. Math. Surveys}, 46, N5:1--51, 1991.

\bibitem{09-6}
J.~Vitoria.
\newblock Mutations vs. seiberg duality.
\newblock {\em Journal of Algebra}, 321:816–828, 2009.

\bibitem{97-6}
A.~V. Zabrodin.
\newblock Hirota's difference equations.
\newblock {\em Theoretical and Mathematical Physics}, 113:1347--1392, 1997.

\end{thebibliography}

\end{document}